\documentclass[12pt,preprint]{aastex}
\shorttitle{Short Title} \shortauthors{Wang \& Fan}

\begin{document}


\title{Systematic errors in the determination of Hubble constant
due to the asphericity and non-isothermality of clusters of galaxies}


\author{Y.-G. Wang and Z.-H. Fan}
\affil{Department of Astronomy, Peking University,
    Beijing 100871, China}
\email{wangyg@bac.pku.edu.cn,fan@bac.pku.edu.cn}


%\altaffiltext{1}{Visiting Astronomer, Cerro Tololo Inter-American Observatory.
%CTIO is operated by AURA, Inc.\ under contract to the National Science
%\altaffiltext{2}{Society of Fellows, Harvard University.}
%\altaffiltext{3}{present address: Center for Astrophysics,
%    60 Garden Street, Cambridge, MA 02138}
%\altaffiltext{4}{Visiting Programmer, Space Telescope Science Institute}
%\altaffiltext{5}{Patron, Alonso's Bar and Grill}


\begin{abstract}
Joint analyses on X-ray and Sunyaev-Zel'dovich (SZ) effect
of a cluster of galaxies can give rise to an estimate
on the angular diameter distance to the cluster. With
the redshift information of the cluster, the Hubble constant
$H_0$ can then be derived. Furthermore, such measurements
on a sample of clusters with a range of redshift can potentially
be used to discriminate different cosmological models.
In this paper, we present statistical studies on the systematic
errors in the determination of $H_0$ due to
the triaxiality and non-isothermality of clusters of galaxies.
Different from many other studies that
assume artificially a specific distribution for the
intracluster gas, we start from the triaxial model of
dark matter halos obtained from numerical simulations.
The distribution of the intracluster gas is then derived under the
assumption of the hydrodynamic equilibrium.
For the equation of state of the intracluster gas, both
the isothermal and the polytropic cases are investigated.
We run Monte Carlo simulations to generate samples
of clusters according to the distributions of their
masses, axial ratios, concentration parameters,
as well as line-of-sight directions. To mimic observations,
the estimation of the Hubble constant
is done by fitting X-ray and SZ profiles of a triaxial
cluster with the isothermal and spherical $\beta$-model.
We find that for a sample of clusters with $M=10^{14}h^{-1}\hbox{ M}_{\odot}$
and $z=0.1$, the value of the estimated $H_0$ is positively biased
with $H_0^{peak}(estimated)\approx 1.05H_0(true)$ and
$H_0^{ave}(estimated)\approx 1.05H_0(true)$ for the isothermal case.
For the polytropic case with $\gamma=1.15$,
the bias is rather large with $H_0^{peak}(estimated)\approx 1.35H_0(true)$
and $H_0^{ave}(estimated)\approx 3H_0(true)$.
For a mass-limited sample of clusters with
$M_{lim}=10^{13}h^{-1}\hbox{ M}_{\odot}$, the results are similar.
On the other hand, such a large overestimation has not been seen
in real observations. It is noticed that the
$\beta$ value for observed clusters is within the
range of $0.5$ to $0.8$. Considering only the subsample of clusters
in Monte Carlo simulations with $\beta$ in the range of $0.5-0.8$,
our analyses show that $H_0^{ave}(estimated)=1.002H_0(true)$
and $H_0^{ave}(estimated)=0.994H_0(true)$ for the
isothermal and polytropic cases, respectively. We further
find that the value of $\beta$ is more sensitive to the
intrinsic asphericity of clusters of galaxies than the axial ratio of
two-dimensional X-ray images $\eta$ is. Limiting to
clusters with $\beta\ge 0.5$ essentially excludes highly
aspherical clusters from the sample. From this subsample of clusters,
we can get a fair estimate on $H_0$.

\end{abstract}

\keywords{distance scale--- galaxies: cluster: general--- dark matter}
%\keywords{cosmology: theory--- galaxy: cluster --- large-scale
%structure of universe}

\section{Introduction}
The measurements of cosmological distances have played
important roles in the development of cosmology.
The discovery of the linear distance-redshift relation
for galaxies in the 1920s (e.g., Hubble 1929) laid the
observational foundation of the big bang cosmological theory.
The accelerating expansion of the universe revealed by
the observations on Type Ia supernovae points to the existence of
dark energy, which has a profound impact on cosmology
as well as on physics (e.g., Perlmutter et al. 1999;
Riess et al. 1998). The commonly used
distance measurements rely on various standard candles, such as
Cepheid variables, Type Ia supernovae, and the Tully-Fisher
relation. The calibration of one particular standard candle often
involves another standard candle. This is the so called distance
ladder. On the other hand, joint analyses of X-ray and the
Sunyaev-Zel'dovich effect of a cluster of galaxies can give
rise to an estimate of the angular diameter
distance to the cluster. This method is independent of the
distance ladder, and therefore provides an important test on
the consistency of cosmological theories.

For clusters of galaxies, their typical mass is
about $10^{14}-10^{15}M_{\odot}$. Besides luminous galaxies,
a large fraction of the baryonic matter
in a cluster is in the form of hot intracluster gas.
With a temperature of a few $keV$, the intracluster gas is
fully ionized. Hot electrons emit strong X-ray through
Bremsstrahlung processes (e.g., Rosati et al. 2002).
Meanwhile, the electrons interact with Cosmic Microwave Background (CMB)
photons and distort the CMB spectrum. The distortion
due to the thermal motion of electrons is referred to as the thermal
Sunyaev-Zel'dovich effect (SZ) (Sunyaev \& Zel'dovich 1970, 1972).
The Bremsstrahlung X-ray emission depends on the density of electrons
through $S_x\propto \int n_e^2\Lambda_{e\mathrm H}(T_e)dl$,
where $n_e$ and $T_e$ are the number density and temperature
of electrons, respectively, $\Lambda_{e \mathrm H}$ is the X-ray
cooling function, and the integration is along the line of sight.
For the SZ effect, we have $\delta T\propto\int n_eT_e{\mathrm d}l$
(e.g., Carlstrom et al. 2002). From dimensional analyses,
one can see that, up to a temperature-dependent
factor, the quantity $(\delta T)^2/S_x$ gives rise to an estimate
of the dimension of the cluster along the line of sight.
Under the assumption of spherical symmetry, this size is the same
as the linear extension of the cluster over the sky.
With its measured angular size,
one can then obtain the angular diameter distance to the cluster,
and further the Hubble constant $H_0$
(McHardy et al. 1990; Birkinshaw et al. 1991; Jones et al.
1993; Birkinshaw \& Hughes 1994; Reese et al. 2002; Reese 2004;
Schmidt et al. 2004).

Observationally, the spherical and isothermal $\beta$ model is
commonly used to describe the distribution of the intracluster gas.
The two parameters in the model, the power index $\beta$ and the
characteristic scale $r_c$, are usually obtained by fitting the
observed X-ray surface brightness with the $\beta$ model.
Thus any deviation from the sphericity
and isothermality for the intracluster gas can introduce significant
biases and uncertainties to the estimation of $H_0$.
Studies have been performed on the effects of non-isothermality
of the temperature profile, and the clumpiness and asphericity of the
intracluster gas distribution (Birkinshaw et al. 1991; Inagaki et al. 1995;
Roettiger et al. 1997; Cooray 1998; Hughes \& Birkinshaw 1998; Sulkanen 1999;
Puy et al. 2000; Udomprasert et al. 2004).
It is noted that the non-isothermality can lead to about $20-30\%$
errors in the $H_0$ determination. The clumpiness and asphericity
may introduce about $15\%$ errors (e.g., Inagaki et al. 1995).
Most of these studies either assume an aspherical gas distribution of
a specific form, e.g., the triaxial $\beta$ model
(e.g.,  Sulkanen 1999),
or concentrate on individual clusters from numerical simulations
(e.g., Inagaki et al. 1995; Roettiger et al. 1997; Ameglio et al. 2005;
Hallman et al. 2005). In this paper, rather than assuming an
ad hoc gas distribution, we investigate the systematic errors
starting from the triaxial mass distribution of
dark matter halos. Our focus is on the statistical analyses for
a sample of clusters generated from Monte Carlo simulations.

From numerical simulations, Jing and Suto (2002) proposed
a triaxial model with specified statistics of the axial ratios
for dark matter halos. The model has been applied to lensing
studies (e.g., Oguri et al. 2003, 2004; Keeton et al. 2004; Tang \& Fan 2005),
to the distribution of intracluster gas (e.g., Lee \& Suto 2003),
and to the gas-associated X-ray and SZ effect of clusters of galaxies
(Lee \& Suto 2004; Wang \& Fan 2004). In our current analyses on
the systematic errors on $H_0$ determination, we adopt the triaxial model
for dark matter halos. The distribution of the intracluster gas
and further the profiles of the X-ray surface brightness and SZ effect
are derived under the assumption of hydrodynamic equilibrium.
We consider both isothermal and polytropic equations of state
for the intracluster gas.
Regarding such distributions as $^{\prime}$true$^{\prime}$ distributions,
the X-ray and SZ effect profiles are fitted with the spherical and isothermal
$\beta$ model. The two parameters $\beta$ and $r_c$
are obtained from the X-ray fitting alone as many observational analyses do.
Then the estimated $H_0$ and the true $H_0$ are compared.
The effects of the non-isothermality and the asphericity,
as well as the effects of some other factors, are analyzed.
To evaluate the biases and uncertainties statistically,
we generate samples of clusters with Monte Carlo simulations.
We take into account the distributions of the
axial ratios and the concentration parameters of dark matter halos.
The mass function of dark matter halos is taken from Jenkins et al. (2001).
Because of the asphericity of clusters, we also need to
consider the statistics of line-of-sight directions
in producing X-ray and SZ effect maps.

The rest of the paper is organized as follows. In \S2, we present the
intracluster gas distribution from the triaxial model of
dark matter halos. In \S3, we describe the
$H_0$ determination from the joint X-ray and SZ analysis.
Our results are shown in \S4 and \S5. \S6 contains conclusions and
discussions.

%Throughout the paper, we consider the concordance cosmological
%model (Spergel et al. 2003) with
%$\Omega_m=0.3=1-\Omega_{\Lambda}=0.7$, $H_0$ = 72 km $\mathrm
%s^{-1}$ $\mathrm{Mpc}^{-1}$ and $\sigma_8=0.9$, where $\Omega_m$
%and $\Omega_{\Lambda}$ are the present dimensionless matter
%density and the dark energy density of the universe, respectively,
%$H_0$ is the present Hubble constant, and $\sigma_8$ is the rms of
%the extrapolated linear mass density fluctuation smoothed over 8
% Mpc$h^{-1}$ with $h$ the Hubble constant in units of 100 km
%$\mathrm s^{-1}$ $\mathrm{Mpc}^{-1}$. Unless specifically stated
%otherwise, the mass and redshift in this paper are
%$M=10^{14}h^{-1}M_{\odot}$, $z=0.1$, respectively.

\section{Density distributions and temperature profiles of the
intracluster gas}
The dark matter distribution of clusters of galaxies is
described by (Jing \& Suto 2002)
\begin{equation}\label{e1}
\frac{\rho(R)}{\rho_{\mathrm{crit}}}=\frac{\delta_c}{(R/R_0)^\alpha(1+R/R_0)^{3-\alpha}}
\end{equation}
where $R=a({x^2}/{a^2}+{y^2}/{b^2}+{z^2}/{c^2})^{1/2}\  (c\leq
b\leq a)$ is the length of the major axis of a density contour,
$R_0$ is a characteristic scale, $\delta_c$ is a dimensionless
characteristic density
contrast and $\rho_\mathrm{crit}$ is the critical density of the
universe. We take $\alpha=1$ for clusters of galaxies.

We assume that the intracluster gas is in hydrodynamic equilibrium
with the gravitational potential. Then the gas distribution
is determined by
\begin{equation}\label{e2}
\frac{1}{\rho_g}\bigtriangledown P_g=-\bigtriangledown \Phi \ ,
\end{equation}
where $P_g$ and $\rho_g$ are the pressure and mass density of the intracluster
gas, respectively, and $\Phi$ is the gravitational potential
from the total mass distribution. The specific gas distribution
depends on its equation of state.
For the isothermal gas, we have
\begin{equation}\label{e4}
\frac{{\rho}_{g}}{{\rho}_{g0}}={\exp{[-{\frac{1}{K}}(\Phi-{\Phi}_0)}]},
\end{equation}
where $K={k_BT_g}/{\mu m_p}$, in which $k_B$, $T_g$,
$\mu$, and $ m_p$ represent the Boltzmann constant, the gas
temperature, the mean molecular weight, and the proton mass,
respectively. The subscript $¡°0¡±$  denotes the corresponding
value at the central position.
For the polytropic gas satisfying the equation of state
$P_g\propto \rho_g^{\gamma}$, its density and temperature distributions
follow
\begin{equation}\label{e6}
\rho_g=\rho_{g0}\bigg [ 1-\frac{1}{K_0}\frac{\gamma-1}{\gamma}
({\Phi}-{\Phi}_0)\bigg ]^{\frac{1}{\gamma-1}},
\end{equation}
\begin{equation}\label{e7}
T_g=T_{g0}\bigg [ 1-\frac{1}{K_0}\frac{\gamma-1}{\gamma}
({\Phi}-{\Phi}_0)\bigg ],
\end{equation}
where $K_0=k_BT_{g0}/\mu m_p$ with $T_{g0}$ being the central
temperature of the gas.

For a cluster of galaxies, the mass fraction of the gas is about $10\%$.
Therefore its gravitational potential is dominantly determined by the
dark matter component. Thus
\begin{equation}
\bigtriangledown^2\Phi=4\pi G\rho \ .
\end{equation}
Given the triaxial mass distribution of the dark matter in eq.(1),
the gravitational potential can be numerically calculated.
Lee and Suto (2003) derived an analytical expression of $\Phi$
under the approximation of small asphericity for dark matter halos.
The perturbative solution can be written in the following form
\begin{eqnarray}\label{e8}
\Phi({\textbf{\emph{u}}}^{\prime})&=&C\bigg\{F_1(u^{\prime})+\frac{e_b^2+e_c^2}{2}F_2(u^\prime) \nonumber\\
&&+\frac{1}{2{r^\prime}^2}\bigg[e_b^2(x^\prime\cos\phi-y^\prime\sin\phi\cos\theta+z^\prime\sin\phi\sin\theta)^2\nonumber\\
&&+e_c^2(y^\prime\sin\theta+z^\prime\cos\theta)^2\bigg]F_3(u^\prime)\bigg\}
\end{eqnarray}
where ${\textbf{\emph{u}}}^{\prime}=
{\textbf{\emph{r}}}^{\prime}/R_0$, ${\textbf{\emph{r}}}^{\prime}
=(x^\prime, y^\prime, z^\prime)$ with $z^\prime$ being the
line-of-sight direction, $\theta$ and $\phi$ are the polar
coordinates of $z^\prime$ in the $(x, y, z)$
coordinate system (Wang \& Fan 2004), $C=4\pi G\delta_c\rho_{crit}{R_0}^2$,
$e_b=(1-{b^2}/{a^2})^{1/2}$ and $e_c=(1-{c^2}/{a^2})^{1/2}$. Here
$e_b$ and $e_c$ are the two eccentricities of the ellipsoidal dark
matter halos. The three functions $F_1(u^\prime)$, $F_2(u^\prime)$
and $F_3(u^\prime)$ are given in Lee and Suto (2003). Wang and Fan (2004)
further derived the X-ray and SZ profiles under the perturbative
approximation, which are the main formulations
to be used in our following analyses.

\section{Estimates on the Hubble constant}
In the observational analyses on X-ray and SZ effect
of a cluster of galaxies, the spherical and isothermal
$\beta$ model is widely adopted to describe
the distribution of the intracluster gas. With this model,
the number density and the temperature of electrons are
\begin{equation}
n_e=n_{e0}[1+(r/r_c)^2]^{-3\beta/2},
\end{equation}
\begin{equation}
T_e=T_{e0},
\end{equation}
where $n_{e0}$ is the central number density of electrons,
$r_c$ is the core radius, and $\beta$ represents the slope
of the profile. The corresponding X-ray and SZ effect profiles are
\begin{equation}
S_x=S_{x0}[1+(\theta/\theta_c)^2]^{1/2-3\beta},
\end{equation}
and
\begin{equation}
\delta T=\delta T_{0}[1+(\theta/\theta_c)^2]^{1/2-3\beta/2}.
\end{equation}
where $\theta=r/D_A$ with $D_A$ being the angular diameter
distance to the cluster.
In the $\beta$ model, the central values $S_{x0}$ and $\delta T_0$
depend on $n_{e0}$, $\theta_c$, $\beta$ and $D_A$.
Observationally, $S_{x0}$ and $\delta T_0$ are measurable quantities,
and $\theta_c$ and $\beta$ are usually obtained through the fitting to the
X-ray profile. Thus by eliminating $n_{e0}$, the angular diameter
distance $D_A$ can be estimated by (e.g., Reese et al. 2002)
\begin{equation}\label{e13}
D_\mathrm{A}(estimated)=\frac{(\delta
T_0)^2}{S_{x0}}(\frac{m_ec^2}{k_BT_{e0}})^2\frac{\Lambda_{eH0}\mu_e/\mu_H}{4\pi^{3/2}
\Re^2_{(x,T_e)}T_{CMB}^2\sigma_T^2
 (1+z)^4}\frac{1}{\theta_c}\frac{1}{\mathrm G(\beta)}
 \end{equation}
where ${\mathrm G(\beta)} = [{\Gamma(3/2\beta-1/2)}/{\Gamma(3/2\beta)}]^2
 {\Gamma(3\beta)}/{\Gamma(3\beta-1/2)}$ with $\Gamma(\beta)$ being
the Gamma function, and $\Re_{(x,T_e)}$ is the frequency-dependent
factor for the SZ effect with $x=h\nu/k_BT_{CMB}$. The quantities
$h$, $k_B$, $c$, $m_e$, $\sigma_T$ and $T_{CMB}$ are
the Planck constant, the Boltzmann constant, the speed of light,
the electron mass, the Thomson cross section, and the CMB temperature,
respectively. The X-ray cooling function is denoted by $\Lambda_{eH0}$ for $T_e=T_{e0}$,
and $n_H=n_e\mu_e/\mu_H$ with $n_j=\rho_g/\mu_j m_p$ for
species $j$. From the functional form of $\mathrm G(\beta)$,
it is noted that eq.(12) can be used only with $\beta>1/3$.

For a cluster with its gas distribution being different from that of the
spherical and isothermal $\beta$ model, the estimated $D_A$ deviates
from the true $D_A$. Thus systematic errors are introduced in
the $H_0$ determination. In our analyses, we treat the gas distribution resulting from the triaxial model of dark matter halos as
the $^{\prime}$true$^{\prime}$ gas distribution.
For the isothermal case, the electron
number density from the triaxial model can be
approximated by (Wang \& Fan 2004)
\begin{eqnarray}\label{e9}
n_e&=&n_{e0}\exp\bigg \{-\frac{C}{K}\bigg\{F_1(u^{\prime})+\frac{e_b^2+e_c^2}{2}F_2(u^\prime) \nonumber\\
&&+\frac{1}{2{r^\prime}^2}\bigg[e_b^2(x^\prime\cos\phi-y^\prime\sin\phi\cos\theta+z^\prime\sin\phi\sin\theta)^2\nonumber\\
&&+e_c^2(y^\prime\sin\theta+z^\prime\cos\theta)^2\bigg]F_3(u^\prime)\bigg\}
\bigg\}.
\end{eqnarray}
The corresponding SZ effect and the X-ray profiles are
\begin{eqnarray}\label{e10}
\delta T&=&\Re_{(x,T_e)}T_{\mathrm{CMB}}D_\mathrm
{A}(true)\theta_0\int\!\sigma_T\frac{k_BT_e}{m_ec^2}n_e
 d(z^\prime/R_0)\nonumber\\
&=&n_{e0}\Re_{(x,T_e)}T_{\mathrm{CMB}}D_\mathrm
{A}(true)\theta_0\sigma_T\frac{K_BT_e}{m_ec^2}f(\theta_x/\theta_0,
\theta_y/\theta_0, c_e, e_b, e_c, \theta, \phi),
\end{eqnarray}
and
\begin{eqnarray}\label{e11}
S_x&=&\frac{1}{4\pi(1+z)^4}D_\mathrm{A}(true)\theta_0\int\!\frac{\mu_e}{\mu_\mathrm
H}n_e^2\Lambda_{e\mathrm H}{d}(z^\prime/R_0)\nonumber\\
&=&{n_{e0}}^2\frac{1}{4\pi(1+z)^4}D_\mathrm{A}(true)\theta_0\frac{\mu_e}{\mu_\mathrm
H}\Lambda_{e\mathrm H} g(\theta_x/\theta_0, \theta_y/\theta_0,
c_e, e_b, e_c, \theta, \phi),
\end{eqnarray}
where $f=1/n_{e0}\int\!n_ed(z^\prime/{R_0})$,
$g= 1/{n_{e0}^2}\int\!n_e^2{d}(z^\prime/R_0)$,
$\theta_0=R_0/D_A(true)$ is the angular scale
radius, and $z$ is the redshift of the cluster.
Then we have
 \begin{equation}\label{e12}
 D_\mathrm{A}(true)=\frac{(\delta
 T_0)^2}{S_{x0}}\bigg(\frac{m_ec^2}{k_BT_{e0}}\bigg)^2\frac{\Lambda_{e\mathrm{H0}}\mu_e/\mu_\mathrm{H}}{4\pi \Re^2_{(x, T_e)}T_{CMB}^2\sigma_T^2
 (1+z)^4}\frac{1}{\theta_0}\frac{g(0, 0, C/K, e_b, e_c,\theta,\phi)}{f^2(0, 0, C/K, e_b, e_c, \theta, \phi)}.
 \end{equation}

Comparing $D_A(true)$ and $D_A(estimated)$, we obtain
\begin{equation}\label{e14}
\frac{D_\mathrm{A}(true)}{D_\mathrm{A}(estimated)}=
\pi^{1/2}\mathrm G(\beta)\frac{\theta_c}{\theta_0} \frac{g(0, 0,
c_e, e_b, e_c,\theta, \phi)}{f^2(0, 0, c_e, e_b, e_c,\theta,
\phi)}.
\end{equation}
Given a cosmological model, we have $D_A\propto H_0^{-1}$, therefore,
\begin{eqnarray}\label{e15}
 E_1&=&\frac{H_0\mathrm{(estimated)}}{H_0\mathrm{(true)}}\nonumber\\
&=&\pi^{1/2}\mathrm G(\beta)\frac{\theta_c}{\theta_0} \frac{g(0,
0, c_e, e_b, e_c, \theta, \phi)}{f^2(0, 0, c_e, e_b, e_c,\theta,
\phi)}.
\end{eqnarray}

Similarly, one can derive $E_2=H_0(estimated)/H_0(true)$ for
the polytropic case, which is given by
\begin{eqnarray}\label{e18}
E_2&=&\frac{H_0(estimated)}{H_0(true)}\nonumber\\
  &=&\pi^{1/2}\mathrm G(\beta)\frac{\theta_c}{\theta_0}
\frac{g^\prime(0, 0, c_e, e_b, e_c, \theta, \phi)}
{{f^\prime}^2(0, 0, c_e, e_b, e_c, \theta,
\phi)}\bigg(\frac{T_\mathrm{ew}}{T_{e0}}\bigg)^{\frac{3}{2}}
\end{eqnarray}
where $T_\mathrm{ew}$ is the emission-weighted temperature,
and the two functions $f^\prime$ and
$g^\prime$ are defined as
\begin{equation}\label{e19}
f^\prime(\theta_x/\theta_0, \theta_y/\theta_0, c_e, e_b, e_c,
\theta,
\phi)=\int\bigg\{1-\frac{1}{K_0}\frac{\gamma-1}{\gamma}[\Phi-\Phi_0]\bigg\}^{\frac{\gamma}{\gamma-1}}
d(z^\prime/R_0)
\end{equation}
\begin{equation}\label{e20}
g^\prime(\theta_x\theta_0, \theta_y\theta_0, c_e, e_b, e_c,
\theta,
\phi)=\int\bigg\{1-\frac{1}{K_0}\frac{\gamma-1}{\gamma}[\Phi-\Phi_0]\bigg\}^{\frac{\gamma+3}{2(\gamma-1)}}
d(z^\prime/R_0)
\end{equation}

It is seen from eq.(18) and eq.(19) that the errors on $H_0$
estimation depend on the mass distribution of clusters through
$\theta_0$, $c_e$, $e_b$, $e_c$, and on the equation of state
of the intracluster gas. Furthermore,
because of the asphericity of a cluster, different observing
directions also give rise to different results. The parameters
$\theta_c$ and $\beta$ are obtained by fitting
the X-ray profile from the triaxial model with the $\beta$-model.
Thus the values of $\theta_c$ and $\beta$ are dependent of
the properties of clusters as well. Besides, because the
$^{\prime}$true$^{\prime}$ X-ray profile for a triaxial cluster does not follow the
$\beta$-profile perfectly, the fitted $\theta_c$ and $\beta$
are affected by the maximum radius to which the fitting is applied
(e.g., Navarro et al. 1995; Komatsu \& Seljak 2001).
It is noticed that $E_1$ and $E_2$ are very sensitive
to the value of $\beta$. Figure 1 shows the function of $G(\beta)$.
It is seen that with $\beta<0.5$, the values of $E_1$ and $E_2$ can be
very large. For the polytropic case, the region used to
obtain the emission weighted temperature $T_{ew}$ also matters
in the estimation of $H_0$.

\section{Results}
In this section, we focus on the effects of the maximum fitting radius,
the temperature gradient, and the polytropic index. We consider
spherical clusters only. In the next section, we present
our statistical analyses taking into account the asphericity
of clusters.

Theoretically, both the spherical NFW model and the
triaxial model assume an infinite extension of the mass
distribution of dark matter halos. In reality however,
an astrophysical object always has a finite size. In our following
analysis, we cut the integration range in eq.(14) and eq.(15)
at $r_{int}=5 r_{vir}$.

\subsection{Effect of the maximum fitting radius}
Because the X-ray emission depends on the square of the number density
of electrons, it decreases sharply toward the outer part of a cluster. Thus
in real observations, the fitting of the $\beta$-model to the
X-ray profile is often performed in the inner part of a
cluster where the signal-to-noise ratio is high enough.
As we explained in the last section, different maximum fitting radii
give rise to different values of $\theta_c$ and $\beta$, and thus
different estimates on the Hubble constant $H_0$.

Figure 2 shows the effects of $r_{fit}$ on $E_1$ and $E_2$
for the isothermal and polytropic cases, respectively.
The polytropic index $\gamma$ is taken to be $1.15$ in the case of
$E_2$.
It is noted that in the isothermal case, the Hubble constant $H_0$
is systematically overestimated with the maximum deviation
occurring at $r_{fit}\sim 0.6 r_{vir}$. For clusters of
$M=10^{14}h^{-1}\hbox{ M}_{\odot}$, the value of
$H_{0}(estimated)$ can deviate from the $H_{0}(true)$ by
as much as $\sim 12\%$. The deviations are larger for smaller clusters.
For the polytropic gas,
when $r_{fit}\leq 0.2r_{vir}$, an overestimation is observed
and the deviations are larger for larger clusters.
With $r_{fit}> 0.2r_{vir}$, the value of $H_0$ is systematically
underestimated with the maximum underestimation of $\sim 12\%$
at $r_{fit}\sim 0.6 r_{vir}$. The smaller the clusters are, the
larger the underestimations are.
In the following analyses, we will take $r_{fit}=0.6 r_{vir}$.
For the NFW mass profile with $\alpha=1$, the mass within
$0.6 r_{vir}$ is about $0.7\hbox{ M}_{vir}$,
and the average density within it is therefore about $3 \rho_{vir}
\approx 300 \rho_{crit}$ at $z=0$ (e.g., Komatsu \& Seljak 2001).
Here $\rho_{vir}=M_{vir}/(4\pi/3 r_{vir}^3)$ and $\rho_{crit}$ is the
critical density of the universe.

\subsection{Effect of the temperature gradient}
When fitting the X-ray profile with the isothermal $\beta$ model,
the temperature used is the emission-weighted temperature $T_{ew}$.
For the polytropic gas with a temperature distribution,
$T_{ew}$ is different from the central temperature $T_{e0}$, and
thus the quantity $(T_{ew}/T_{e0})^{3/2}$ in eq. (19) contributes
an additional factor to $E_2$. It is clearly seen that $T_{ew}$
depends on the temperature profile as well as on the region
where the emission-weighted temperature is derived.

Given the temperature profile in the polytropic gas, we can
calculate the emission-weighted temperature through the following
equation (e.g., Ascasibar et al. 2003; Rasia et al. 2005)
\begin{equation}
T_{\mathrm{ew}}\equiv\frac{\int{n_e}^2T^{3/2}dV}
{\int{n_e}^2T^{1/2}dV}.
\end{equation}
%It is known that the distribution of temperature profile will be
%equilibrium (Lee \& Suto 2003). In order to illustrate the effect
%of the $T_\mathrm{ew}$ in the $H_0$ estimation, we only consider
%the simplest case of the distribution of temperature profile
%$$(e_b=0, e_c=0)$.

For a spherically symmetric gas distribution, the emission-weighted
temperature can be written as
\begin{equation}
T_{\mathrm{ew}}\equiv\frac{\int_0^{r_T}{n_e}^2T^{3/2}r^2dr}
{\int_0^{r_T}{n_e}^2T^{1/2}r^2dr}.
\end{equation}
Obviously, $T_{ew}$ depends on the
integral range $r_\mathrm{T}$. Figure 3 shows the influence
of $r_\mathrm{T}$ on the $H_0$ estimation.
In our calculations, we take $\gamma=1.15$, and the $\beta$ profile fitting is done
out to a fixed radius of $r_{fit}=0.6r_{vir}$ when $r_\mathrm{T}$
is varying. When $r_\mathrm{T}$ is small, $T_{ew}/T_{e0}$ is close
to unity, and $E_2>1$. With $r_\mathrm{T}>0.2r_{vir}$, $E_2<1$
and reaches $\sim 0.9$ as $r_\mathrm{T} \rightarrow r_{vir}$.
From Figure 3, it is seen that if we ignore the difference between
$T_{ew}$ and $T_0$, i.e., take $T_{ew}/T_{e0}=1$, $E_2>1$.
Thus the underestimation appeared in the polytropic case is mainly
due to $T_{ew}/T_{e0}<1$.
In the analyses presented in Figure 2 as well as in the following studies,
we take $r_\mathrm{T}=0.6r_{vir}$.

\subsection{The effect of the polytropic index}
Given the mass distribution of a dark matter halo, both the
density and the temperature profiles depend on the polytropic
index $\gamma$. In this subsection, we discuss how $\gamma$
affects the $H_0$ estimation.

Figure 4 presents the dependence of $E_2$ on $\gamma$.
The fitting radii for the $\beta$ profile and the emission-weighted
temperature are $r_{fit}=0.6r_{vir}$ and $r_T=0.6r_{vir}$, respectively.
The mass of the cluster is $M=10^{14}h^{-1}\hbox{ M}_{\odot}$.
The result shows that $E_2<1$ when $1.05\le \gamma \le 1.4$, and $E_2>1$
when $\gamma > 1.4$. With $\gamma$ approaching $1$, the gas
becomes isothermal and $E_2$ reaches $\sim 1.12$, the result of
$E_1$ in Figure 2. At large $\gamma$ with $\gamma\sim 2$, the value
of $E_2$ can be as high as about $1.6$. We note that in the most theoretically
probable range of $\gamma\sim 1.15$ for the intracluster gas
(Komatsu \& Seljak 2001), the $H_0$ is underestimated with
the level of about $10\%$. Our results here are consistent
with those of Puy et al. (2000) and Udomprasert et al. (2004)

Observationally, it is difficult to determine $\gamma$ precisely.
The available values of $\gamma$ for clusters of galaxies are around $\gamma \sim 1.2$
(e.g., Hughes et al. 1988; Finoguenov, Reiprich \& Bohringer 2001;
Pratt, Bohringer \& Finoguenov 2005).
Thus the polytropic index can be one of the factors that lead
to the underestimation of $H_0$ appeared in most of the observations.

%To a real galaxy cluster, once the density and temperature
%profiles are given, then the polytropic index $\gamma$ can be
%obtained by fitting the temperature profile (Hughes et al. 1988).
%Nevertheless, in the theoretical prediction, the case is just
%inverse: only given the polytropic index $\gamma$ firstly, we can
%have the density and temperature profiles. Although the
%theoretical prediction (Komatsu \& Seljak 2001), the numerical
%simulation (Ascasibar et al. 2003) and observation (Finoguenov et
%al. 2001) can give the distribution of the polytropic index
%$\gamma$, it is still tough to give an accurate distribution of
%polytropic index for the kinds of uncertainty factors (merging,
%cooling flow etc) in galaxy clusters. So, we adopt $\gamma=1.15$
%in this paper unless specifically stated.
%
% Figure \ref{fig5} illustrates the relative
%error for the Hubble constant between a isothermal $\beta$-model
%and polytropic model. For a relatively small $\gamma\sim1.2$, it
%could be underestimated by $10\%$; while to a large $\gamma$, it
%may be overestimated nearly by a factor of two. Our results are
%similar to Puy et al. (2000) and Udomprasert et al. (2004) except
%the quite smaller polytropic index($\gamma<1.25$). For the small
%polytropic index, the distribution of gas approaches
%isothermality, however, note that even using $\beta$-model to fit
%the X-ray surface brightness profile which is predicted by the
%isothermal gas in hydrostatic equilibrium with the spherical dark
%halo can lead to an overestimation of $H_0$ by $8\sim11\%$.

\section{Statistical analyses on the $H_0$ estimation in triaxial clusters}
In this section, we discuss statistically the effects of the
triaxiality of clusters on the $H_0$ determination.
Monte Carlo simulations are employed in the studies.

From eq.(18) and eq.(19), it is known that $E_1$ and $E_2$
depend on the mass distribution of dark matter halos through
the concentration parameter $c_e$, and the eccentricities
$e_b$ and $e_c$ (or equivalently the axial ratios $b/a$ and $c/a$),
as well as on the observing direction $(\theta, \phi)$.
It should be noted that the values of $\beta, \theta_c$, and
$T_{ew}/T_{e0}$ are also functions of
$(c_e, e_b, e_c, \theta, \phi)$.

We first analyze the effects of line-of-sight directions.
Given the mass configuration of a cluster, the
probability function of $E_1$ can be written as
\begin{equation}\label{e22}
F\bigg(E_1\bigg|c_e, \frac{c}{b}, \frac{c}{a}\bigg)={2\over
\pi}\int_{\theta_1}^{\theta_2} \bigg (\frac{\partial
E_1}{\partial\phi}\bigg )^{-1}_{\theta, c_e, c/b,
c/a}\sin{\theta}d\theta,
\end{equation}
where $[\theta_1,\theta_2]$ is the integral range of $\theta$ within
which a physically meaningful value of $\phi$ can be found for a
given value of $E_1$ (Binney \& de Vaucouleurs 1981). The
distribution function of $E_2$ has the similar form.
The results are presented in Figure 5. The top and bottom panels are for the
isothermal and polytropic cases, respectively. The position of
the vertical line in each plot indicates the peak value of $E_1(E_2)$.
Two sets of $(e_b, e_c)$ are considered. They are $(e_b, e_c)=(0.2, 0.3)$
and $(0.6, 0.8)$. The corresponding axial ratios are $(b/a, c/a)=(0.98, 0.95)$
and $(0.8, 0.6)$. In all the cases, the concentration parameter is taken to
be the average one calculated from Jing and Suto (2002). The mass of the
cluster is $M=10^{14}h^{-1}\hbox{ M}_{\odot}$. The redshift is $z=0.1$.
For the polytropic case, the value of $\gamma$ is taken to be $1.15$.

With small $e_b$ and $e_c$ (left panels), the distribution functions
of $E_1$ and $E_2$ are peaked at $\sim 1.1$ and $\sim 0.9$
with narrow extensions, respectively. For the isothermal case,
the distribution concentrates on the side of $E_1>1$.
For the polytropic case, the distribution is on the side of $E_2<1$.
Because of the small asphericity, the results are similar to those of
the spherical cases presented
in Figure 2 with $r_{fit}=0.6r_{vir}$. It is known that
the combined analysis of X-ray and SZ effects of a cluster
leads directly to an estimate on the line-of-sight dimension
of the cluster. With an observed angular extension of the cluster,
the angular diameter distance to the cluster is determined
by assuming the equality of the sizes of the cluster parallel and
perpendicular to the line of sight. Thus for a triaxial
cluster oriented with its long axis
along the line of sight, one may expect an overestimation for
the angular diameter distance, and therefore an underestimation for
$H_0$. The reverse trend is expected for a cluster with
its short axis along the line of sight. Thus it seems that
the distribution should extend to both $E_1(E_2) >1$ and $E_1(E_2) <1$,
which is not seen in the left panels of Figure 5.
The reason for this is that the above qualitative analysis
considers only the effect of the asphericity.
Taking into account the errors introduced by
the differences between the real gas distribution and that of the
isothermal $\beta$ model, the peak values of $E_1$ and $E_2$ shift away from unity.

With the increase of the asphericity, the distributions become broader.
For the polytropic case with $(e_b, e_c)=(0.6, 0.8)$,
the values of $E_2$ move to the side of $E_2>1$.
The peak value of $E_2$ is around $1.8$.
This large overestimation is mainly due to the small $\beta$
value obtained by fitting the strongly triaxial and polytropic
gas distribution to the isothermal $\beta$ model.

Now the distributions of $E_1$ and $E_2$ are investigated taking into
the statistics of $c_e$, $e_b$ and $e_c$. For a sample of clusters
with a given mass $M$ and at a redshift $z$, we have, for the isothermal
case,
\begin{equation}\label{e21}
F(E_1|{M,z})dE_1=\bigg
[\frac{2}{\pi}\int\!p\bigg(\frac{c}{a}\bigg)d\bigg({c\over
a}\bigg)\int\! p\bigg({c\over b}\bigg|{c\over
a}\bigg)d\bigg({c\over b}\bigg)\int\!p(c_e)dc_e\int\!\bigg
({\partial{E_1}\over \partial {\phi}}\bigg
)^{-1}_{\theta,c_e,c/b,c/a}\sin{\theta}d\theta \bigg ]dE_1,
\end{equation}
where the distributions of $c_e$, $c/b$ and $c/a$ are taken from
Jing and Suto (2002) (Wang \& Fan 2004). For the polytropic case,
the distribution of $E_2$ has the similar form.

Figure 6 shows the results for different $\gamma$. The cluster mass
$M=10^{14}h^{-1}\hbox{ M}_{\odot}$ and $z=0.1$. A clear trend is
observed. As the value of $\gamma$ increases, the distribution
extends more toward high $E_2$. With $\gamma=1.2$, the tail
goes well beyond $E_2=3$. The position of the vertical line in each panel
indicates the peak value of $E_1(E_2)$. The symbols represent
the average values of $E_1(E_2)$. In Figure 7, we
show explicitly the change of the average value of $E_2$ with
$\gamma$. For the isothermal case with $\gamma=1$,
$E_1^{ave}\sim 1.06$. For the
case with $\gamma=1.15$, $E_2^{ave}\sim 3$. It is reminded that
our previous analysis on the spherical clusters with
$\gamma=1.15$ gives the result of $E_2\sim 0.9$ (Figure 2).
Thus the asphericity strongly influences the determination of $H_0$
in the polytropic case due to the small fitted value of $\beta$.

We present the results for clusters with different $M$ at different $z$
in Figure 8. The positions of the vertical lines represent the peak values
of $E_1(E_2)$. The numbers in the parentheses are the average values of
$E_1(E_2)$. For the isothermal case with $z=0.1$, the average value of $E_1$
is $1.05, 1.06$ and $1.08$ for $M=10^{13}h^{-1}\hbox{ M}_{\odot}$,
$10^{14}h^{-1}\hbox{ M}_{\odot}$, and $10^{15}h^{-1}\hbox{ M}_{\odot}$,
respectively. For $M=10^{14}h^{-1}\hbox{ M}_{\odot}$, the values of $E_1^{ave}$
at $z=0.1,0.5$ and $0.9$ are respectively $1.06, 1.08$ and $1.2$.
In these distributions, the peak value of $E_1$ is around $1.05$.
In the case of polytropic gas with $\gamma=1.15$ and $z=0.1$,
we have $E_2^{ave}= 2.6, 3.0$ and $3.5$ for
$M=10^{13}h^{-1}\hbox{ M}_{\odot}$,
$10^{14}h^{-1}\hbox{ M}_{\odot}$, and $10^{15}h^{-1}\hbox{ M}_{\odot}$,
respectively. The corresponding peak values are
$1.05, 1.35$, and $1.35$.
With a fixed mass of $M=10^{14}h^{-1}\hbox{ M}_{\odot}$,
the peak values of $E_2$ are $1.35, 1.55$, and $1.55$
for $z=0.1, 0.5$ and $0.9$, respectively.
The respective average values of $E_2$ are $3.0, 3.6$ and $6.1$.
Thus for massive clusters at relatively high redshifts,
the errors in $H_0$ determination are large due to their strong
asphericities. Observationally, however, high redshift clusters with
strongly distorted X-ray/SZ images are usually not used
in the determination of the angular diameter distance
because very likely they are still in the merging stage.

We further study statistically the errors for a mass-limited cluster
sample with $M_{lim}=10^{13}h^{-1}\hbox{ M}_{\odot}$. The
probability function for $E_1$ ($E_2$) has the following form
(Wang \& Fan 2004)
\begin{equation}
F(E_1)={\int dV(z)\int_{M_\mathrm{lim}} F(E_1|{M,z})(dn/ dM )dM
\over \int (dn/ dM) dM dV},
\end{equation}
where $F(E_1|{M,z})$ is the probability density of $E_1$ for
clusters of mass $M$ and at redshift $z$,
$(dn/dM)dM$ is the number density of clusters in the mass range of $(M,M+dM)$,
and $dV$ is the volume element. The redshift range is taken to be $z=[0, 1]$.
The number density of clusters is modeled based on the
analyses of Jenkins et al. (2001). The results are shown in Figure 9.
For the isothermal case, the distribution
is peaked at $E_1\sim 1$ and $E_1^{ave}\approx 1.06$ with a
dispersion about $\pm 20\%$. For the polytropic case
with $\gamma=1.15$, the distribution function has a peak at
$E_2\approx 1.37$ and a long tail extending to high $E_2$ values.
The average value of $E_2$ is $E_2^{ave}\approx 3$. Thus for a
mass-limited sample of clusters, the $H_0$ estimation
can be highly biased due to the combined effects of non-isothermalities
and asphericities.

There have been different studies on the systematics
involved in $H_0$ determination (e.g., Sulkanen 1999;
Udomprasert et al. 2004). With numerical simulations,
two groups recently analyzed possible errors in estimating $H_0$
(Ameglio et al. 2005; Hallman et al. 2005). While both of them noted the
significant effects of the non-isothermality,
they found opposite tendencies in the $H_0$ determination.
Ameglio et al. (2005) presented an overestimation
on $H_0$, while an underestimation on $H_0$ was reported in
Hallman et al. (2005). Our analyses for the spherical case
show a $\sim 15\%$ underestimation in $H_0$ with
$\gamma$ in the range of $1.05$ to $1.4$. The bias becomes
positive for larger $\gamma$ (see Figure 4). Including
the triaxiality of clusters, however, we find a very large
bias toward overestimation in $H_0$ with $E_2^{ave}\sim 3$
for $\gamma=1.15$. The differences in the results from different groups
are associated with different gas distributions and temperature
profiles adopted in the analyses. While both Ameglio et al. (2005)
and Hallman et al. (2005) rely on numerical simulations,
we derive the properties of the intracluster gas from the
triaxial mass distribution of dark matter halos under
the approximation of hydrodynamic equilibrium. Even
both from simulations, the obtained mass and temperature distributions
of the intracluster gas can be quite different from
one group to another because of the differences
existed in modeling processes. It is also noted that in Ameglio et al. (2005),
just as we have done in this paper, the parameters
$\beta$ and $\theta_c$ were determined by fitting the
$\beta$ model to the X-ray images alone.
Hallman et al. (2005), on the other hand, obtained
the fitting parameters from X-ray and SZ images. This difference
in the fitting technique can be an important reason to account for the
different results shown by the two groups.
The large dispersion appeared in different analyses indicates that the error
analysis itself is model dependent.
Different analyzing procedures can also lead to different
conclusions. In our analyses, each time we change the properties of
the intracluster gas, we obtain a new set of fitting values
for $\beta$ and $\theta_c$. Some other studies may fix the values of
$\beta$ and $\theta_c$ while changing the intrinsic gas properties
(e.g., Udomprasert et al. 2004).

Confronted with observations, the large positive bias
shown in the above studies for the polytropic gas has not been seen
in observational analyses. Most of the results on $H_0$ derived
from joint X-ray and SZ observations are biased low in comparison
with that determined from other observations (e.g., Reese et al. 2002;
Freedman et al. 2001), although for a couple of clusters
the estimated values of $H_0$ are as high as about
$100\hbox{ km/s/Mpc}$ (e.g., Mason et al. 2001, Jones et al. 2003).
There are several issues that might be related to this apparent
discrepancy.

In our studies, we assume that there are no correlations
between the polytropic index $\gamma$ and the triaxiality of
clusters $(e_b, e_c$). Thus, for each $\gamma$,
the analyses are performed for a full sample of clusters
with the statistics of the triaxiality given by Jing and Suto (2002).
It is unclear, however, if $\gamma$ and $(e_b, e_c)$ are
associated with each other. In Figure 10, we
plot the measured $\gamma$ and axial ratios $\eta$ of the
X-ray profiles for clusters of galaxies collected
from literatures. Keeping in mind the large error bars, we see a
tendency that high $\gamma$ is likely related with small $\eta$,
i.e., clusters with higher $\gamma$ tend to be more spherical.
If the correlations indeed exist, the
bias for polytropic and triaxial clusters can be overestimated
(comparing Figure 7 with Figure 4). With numerical simulations,
it is readily testable if there are any relations between
$\gamma$ and the shape of the gas distribution.

Another issue is related to the $\beta$ value, to which
the bias is very sensitive. It is noted that for observed clusters,
most of the $\beta$ values are in the range of $[0.5, 0.8]$.
Considering only the subsample of clusters with $\beta=[0.5, 0.8]$,
our analysis gives $E_1\approx 1.002$ and $E_2\approx 0.994$.

Observationally, clusters with strongly aspherical X-ray/SZ profiles
are often avoided to be used in the $H_0$ measurement.
Here we study the effects of these highly elongated clusters
on the statistical result of $H_0$ for the sample of clusters with
$M=10^{14}h^{-1}\hbox{ M}_{\odot}$, $z=0.1$ and $\gamma=1.15$.
%Constraining us to the subsample of clusters with the axial ratio
%of X-ray surface brightness $\eta \ge 0.95$, we get the distribution
%of $E_2$ with $E_2^{peak}=1.25$ and $E_2^{ave}=2.52$.
In Figure 11, we show the distributions of $E_2$ for
the full sample (solid line), for the subsample with
$\eta \ge 0.95$ (dotted line) and for the subsample with
$\beta\ge 0.5$ (dash-dotted line). Here $\eta$ is the axial ratio
of the X-ray isophote contour at the virial radius of a cluster.
It is seen that the distribution of the subsample with
$\eta \ge 0.95$ indeed shifts toward small $E_2$ but not
significantly in comparison with that of the full sample.
We have $E_2^{peak}(\eta\ge 0.95)\approx 1.25$ and
$E_2^{ave}(\eta\ge 0.95)\approx 2.5$. For the
full sample, the values are $E_2^{peak}\approx 1.35$ and
$E_2^{ave}\approx 3.0$. For the subsample with $\beta\ge 0.5$,
however, the distribution of $E_2$ is narrowly around $E_2\sim 1$.
We therefore immediately see that $E_2$ depends dominantly
on the value of $\beta$, and the long tail in the distribution
of the full sample is resulted from small values of $\beta$.
On the other hand, limiting $\eta$ cannot significantly
affect the $E_2$ distribution.
The apparent reason for this is that for a given $\eta$, a
range of $\beta$ values can exist as shown in Figure 12,
where we apply a cut on $\beta\ge 0.33$.
For each $\eta$, the maximum $\beta$ value
is highly correlated with the value of $\eta$, but the scatter
plot shows a large extension toward small $\beta$. The
existence of these small $\beta$ values results the long tail
in the $E_2$ distribution for the subsample with $\eta\ge 0.95$.
We further analyze the correlations of $\eta$ and $\beta$
with the intrinsic asphericity of dark matter halos represented
by $e_b$ and $e_c$. In Figure 13, we plot the
distributions of $e_b$ and $e_c$ for the three samples respectively.
Clusters in the subsample with $\beta\ge 0.5$ are apparently
biased toward spherical ones in comparison with the full sample.
For the subsample with $\eta\ge 0.95$, however, the distributions
of $e_b$ and $e_c$ shift only slightly to smaller values of $e_b$
and $e_c$. Thus the value of $\beta$ is more
sensitive to $e_b$ and $e_c$ than $\eta$ is.
In Wang and Fan (2004), we analyze the dependence of $\eta$ on
the intrinsic mass distribution of dark matter halos.
Because the distribution of the intracluster
gas follows directly the profile of the gravitational potential,
it is more spherical than the distribution of the dark matter is.
Moreover, the shapes of X-ray/SZ images also depend on
line-of-sight directions. The resulting $\eta$ for the full sample
of clusters discussed here is peaked around $\eta\sim 0.9$.
In other words, while clusters with highly distorted X-ray/SZ images
(small $\eta$) are strongly aspherical intrinsically,
samples with nearly circular images (large $\eta$) can still
contain clusters with high values of $(e_b, e_c)$.
On the other hand, we find that the
radial distribution of the intracluster gas
is very sensitive to the intrinsic asphericity of the dark
matter distribution, and thus the fitted $\beta$ value is
sensitive to $(e_b,e_c)$. Clusters with larger values of $(e_b,e_c)$ tend to
have flatter distributions of the intracluster gas, and
therefore smaller $\beta$ values. In Figure 14,
we show the circularized radial
profiles of the X-ray surface brightness for three specific
clusters all with $\eta=0.95$. Fitting the profiles
with the isothermal $\beta$-model, we get $\beta=0.4,0.5$, and
$0.6$, respectively. The respective $(e_b,e_c)$ of underlying
dark matter distributions are $(0.67,0.77)$, $(0.48,0.71)$ and
$(0.40,0.54)$.
%The examples clearly demonstrate that the
%distributions of the intracluster gas
%for clusters with larger $(e_b, e_c)$ decrease more slowly
%at large radius than those of clusters with smaller $(e_b, e_c)$.
%This slow decrease results a small value of $\beta$.
Combining the results shown in Figure 11 and Figure 13,
we conclude that by considering intrinsically nearly
spherical clusters only, we eliminate the large overestimates on
$H_0$. The $\beta$ value is a better indicator of the
asphericity of underlying dark matter distribution than
the two-dimensional axial ratio of X-ray/SZ isophote contours
$\eta$ is. Our study presented here also opens
a possibility of using the distribution of the $\beta$ value to
probe statistically the intrinsic mass distribution of
dark matter halos.

There are other physical factors that can affect the $H_0$
determination considerably. Analyses based on numerical simulations
show that (e.g., Ameglio et al. 2005)
excluding unrelaxed clusters with apparent substructures from
a sample of clusters can significantly reduce the errors in
the $H_0$ estimation. The intrinsic anisotropies
of CMB can also introduce a notable bias on the $H_0$ determination
(e.g., Udomprasert et al. 2004).

\section{Summary and Discussion}
Combined analyses of X-ray and SZ effect observations provide an
important means in determining the angular diameter distances
to clusters of galaxies. In comparison with the method using
standard candles, this determination does not depend on distance
ladders, and thus avoids the error propagation from
ladder to ladder. Furthermore, the X-ray/SZ observations
and observations on standard candles involve different
systematic errors. Therefore important tests on the consistency
of cosmological theories can be possible by
comparing the results on the evolution of the universe
from two sets of observations.
Besides the determination on $H_0$,
observations on high-redshift clusters can
lead to a Hubble diagram of $D_A-z$ relation, which
potentially can be used to constrain cosmological parameters
just as SNe Ia observations do. In order to extract
useful cosmological information, however, one needs to understand
thoroughly the systematic errors involved in the method
of X-ray/SZ observations.

In this paper, we focus on the triaxiality and
non-isothermality of clusters of galaxies.
From the triaxial model of dark matter halos (Jing \& Suto 2002),
the equilibrium distribution of the intracluster gas is derived,
which is generally aspherical and non-isothermal.
The systematic errors in $H_0$ estimation are
evaluated with Monte Carlo simulations.

Our main results can be summarized as follows:

(i) Different extension adopted in fitting the $\beta$-model
to the profile of X-ray emissions can introduce
as much as about $12\%$ bias in the estimated $H_0$ value for $M=10^{14}h^{-1}\hbox{ M}_{\odot}$.
For the isothermal case, the bias is an overestimate,
while it is an underestimate for the polytropic gas with
$r_{fit}>0.3 r_{vir}$.

%(ii)The finite integral radius can overestimate $H_0$ by almost
%$50\%$ for the quite large eccentricities, such as $e_b=0.6,
%e_c=0.8$ with  the polytropic gas distribution. To the larger
%eccentricities, the error will be larger.

(ii) The maximum radius $r_T$ used to calculate the emission-weighted
temperature in the polytropic case can affect $H_0$ considerably.
With $r_T\leq 0.2r_\mathrm{vir}$, the value of $H_0$ is overestimated.
At larger $r_T$, it is underestimated by about $10\%$ with
$\gamma=1.15$.
%Inagaki et al. (1995) also find this effect in the
%galaxy clusters which have a similar temperature profiles to Coma
%cluster.

(iii) In the spherical case, the effect of the non-isothermality
depends on the polytropic index $\gamma$.
With $1.05\leq \gamma\leq 1.4$, $H_0$ can be underestimated
by nearly $20\%$. A larger $\gamma$ gives rise to a positive bias.
As $\gamma$ is close to the adiabatic value of $1.7$, the overestimate
is about $10\%$.

(iv) With the increase of the asphericity of clusters, the
scatters in the estimated $H_0$ become large. Meanwhile,
the peak value of $H_0$ changes significantly. For the
polytropic case with $\gamma=1.15$, the peak $H_0$ changes
from $H_0^{peak}(estimated)\approx 0.9H_0(true)$ to
$H_0^{peak}(estimated)\approx 1.8 H_0(true)$ as $(e_b, e_c)$
increases from $(0.2,0.3)$ to $(0.6, 0.8)$. The average
$H_0^{ave}(estimated)$ for a sample of clusters with
a given mass $M$ and at a redshift $z$ is very sensitive
to $\gamma$. For $M=10^{14}h^{-1}\hbox{ M}_{\odot}$ and $z=0.1$,
$H_0^{ave}(estimated)\approx 1.7H_0(true)$ at $\gamma=1.05$,
and at $\gamma=1.2$, $H_0^{ave}(estimated)\approx 3.6 H_0(true)$.
Because the asphericities of clusters increase with their masses
and redshifts (Jing \& Suto 2002), the bias on $H_0$ increases
with $M$ and $z$ as well.

(v) For a mass-limited sample of clusters with $M=10^{13}h^{-1}\hbox{ M}_{\odot}$,
we have $H_0^{ave}(estimated)\approx 1.06H_0(true)$ and the scatters
are around $\pm 20\%$ for the isothermal case. For the
polytropic case with $\gamma=1.15$, the peak value of
$H_0^{peak}(estimated)$ is about $1.4H_0(true)$. There is a
very extended tail toward high $H_0(estimated)$, which results
an average value of $H_0^{ave}(estimated)\approx 3 H_0(true)$.

(vi) For $M=10^{14}h^{-1}\hbox{ M}_{\odot}$, $z=0.1$, and $\gamma=1.15$,
we analyze statistically the errors for different subsamples
of clusters. For the subsample of clusters
with $\beta$ being in the range of $[0.5,0.8]$, which is
the range for most observed clusters,
the average value of the estimated $H_0$ is
about $1.002H_0(true)$ and $0.994H_0(true)$ for the
isothermal and polytropic cases, respectively. We also
consider a subsample of clusters with nearly circular X-ray images.
Specifically, we select clusters with the axial ratio
of two-dimensional X-ray images $\eta\ge 0.95$. The distribution
of $E_2$ for this subsample of clusters shifts to smaller $E_2$ side
in comparison with the full sample. But there is still a long tail
toward high values of $E_2$. On the other hand, for clusters with
$\beta\ge 0.5$, the estimated $E_2$ are around $E_2\sim 1$ with
$\pm 10\%$ scatters. We further show that the $\beta$-limited sample
is clearly biased more to intrinsically spherical ones than
the $\eta$-limited sample is. Thus by limiting the value of
$\beta\ge 0.5$, we essentially select clusters that are
nearly spherical intrinsically. For these clusters, we can
obtain a fair estimate on $H_0$ through joint X-ray/SZ analyses.

Based on the triaxial model of dark matter halos, our analyses
avoid ad hoc assumptions on the aspherical distribution of the
intracluster gas. On the other hand, the condition of the
hydrodynamic equilibrium used in our study is likely a
simplification. The real gas distribution and the temperature
profile of a cluster of galaxies can be more complex than
our model can describe (e.g., Kempner et al. 2002;
O'Hara et al. 2004; Reiprich et al. 2004; Inagaki et al. 1995;
De Grandi \& Molendi 2002). With dramatic advances in both
X-ray and SZ observations, however, the knowledge on the
intracluster gas will improve greatly. It is
becoming possible observationally to describe the intracluster
gas more precisely than the isothermal $\beta$-model does.
Therefore better determinations on $H_0$ and even other
cosmological parameters are highly expected from
analyses on clusters of galaxies.

\acknowledgments We are very grateful for the referee's
encouraging and constructive comments and suggestions.
This research was supported in part by the
National Science Foundation of China under grants 10243006 and
10373001, by the Ministry of Science and Technology of China
under grant TG1999075401, by the Key Grant
Project of Chinese Ministry of Education (No. 305001),
and by the National Science Foundation of China under grant 10533010.

\begin{figure}
\plotone{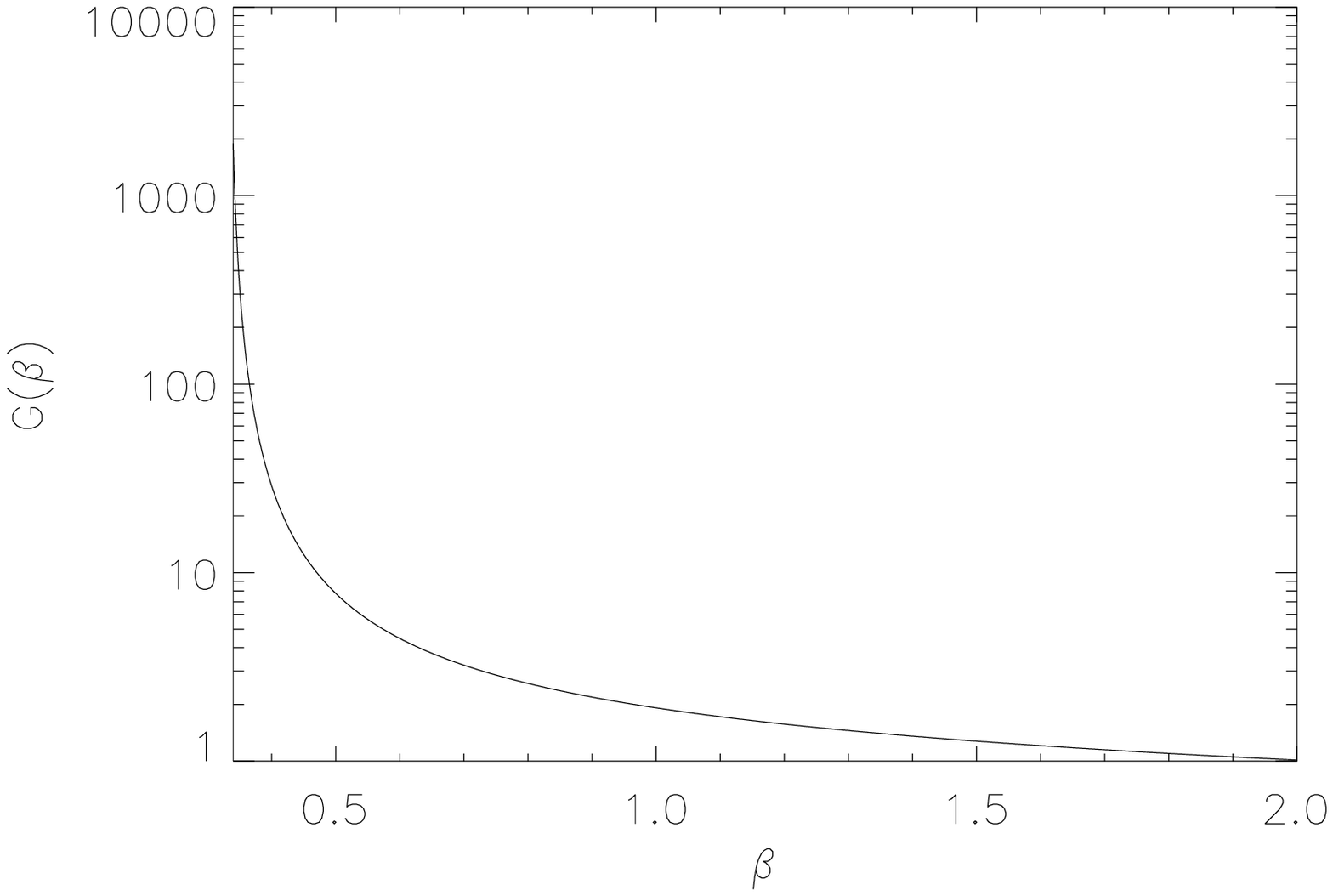}\caption{The dependence of $G(\beta)$ on $\beta$.
\label{fig1}}
\end{figure}

\begin{figure}
\plotone{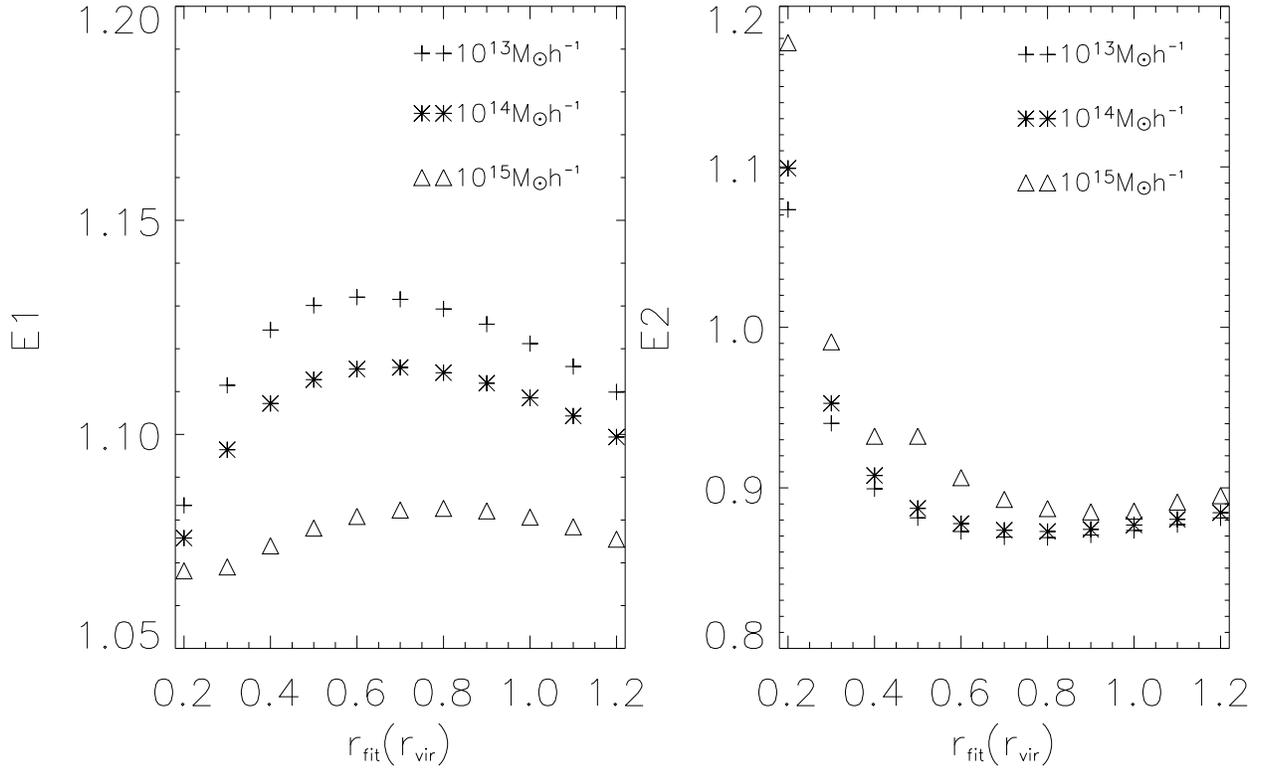} \caption{The dependence of $E_1$ and $E_2$
on the maximum fitting radius $r_{fit}$. Spherical clusters are
considered. The left panel is for the isothermal case,
and the right panel is the polytropic case with $\gamma=1.15$.
The pluses, asterisks and triangles are for $M=10^{13}h^{-1}M_{\odot}$,
$10^{14}h^{-1}M_{\odot}$ and $10^{15}h^{-1}M_{\odot}$, respectively.
\label{fig2}}
\end{figure}

\begin{figure}
\plotone{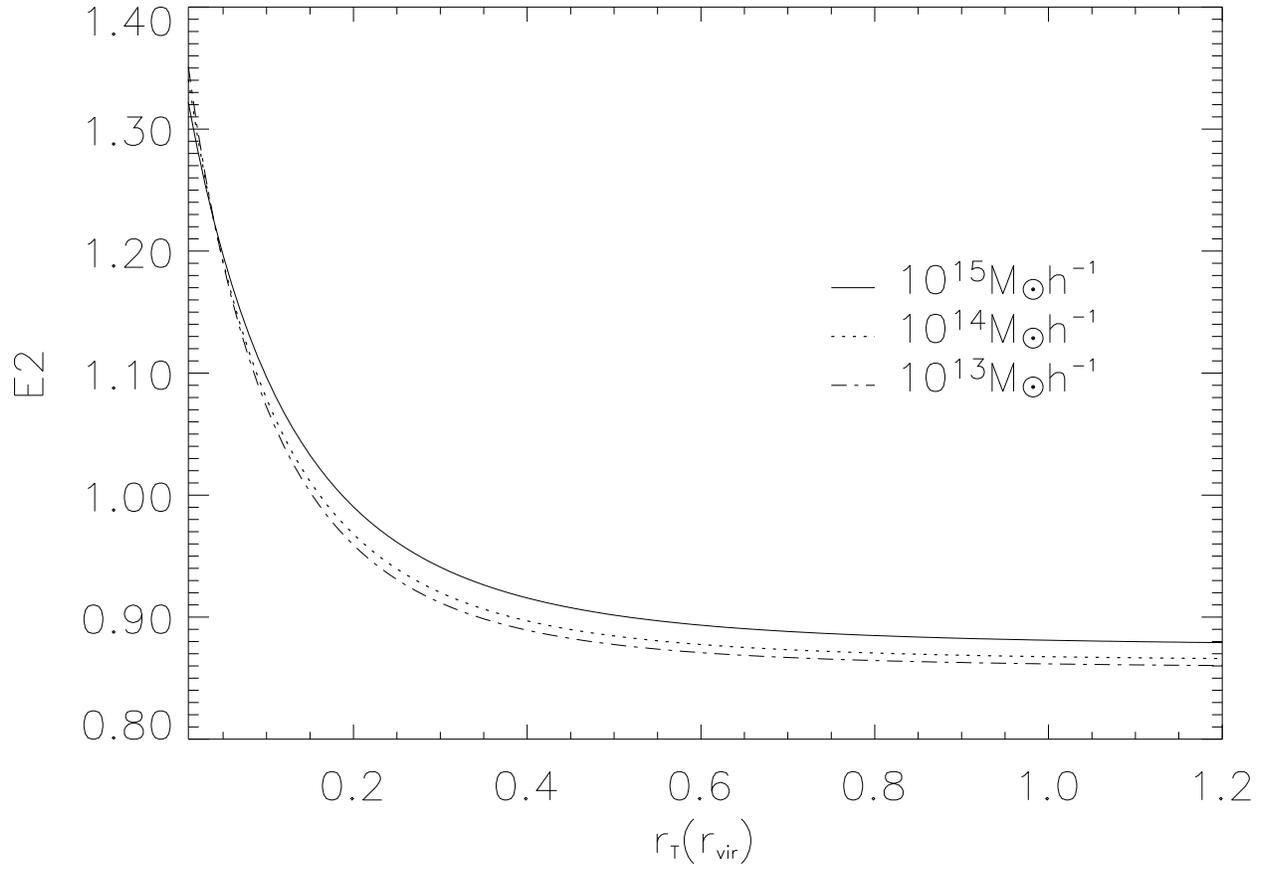}\caption{Influence of $r_T$ on $E_2$.
Spherical clusters are considered, and the polytropic index
$\gamma$ is taken to be $1.15$. The solid, dotted, and dash-dotted lines
are for $10^{15}h^{-1}M_\odot$, $10^{14}h^{-1}M_\odot$, and
$10^{13}h^{-1}M_\odot$, respectively.
\label{fig3}}
\end{figure}

\begin{figure}
\plotone{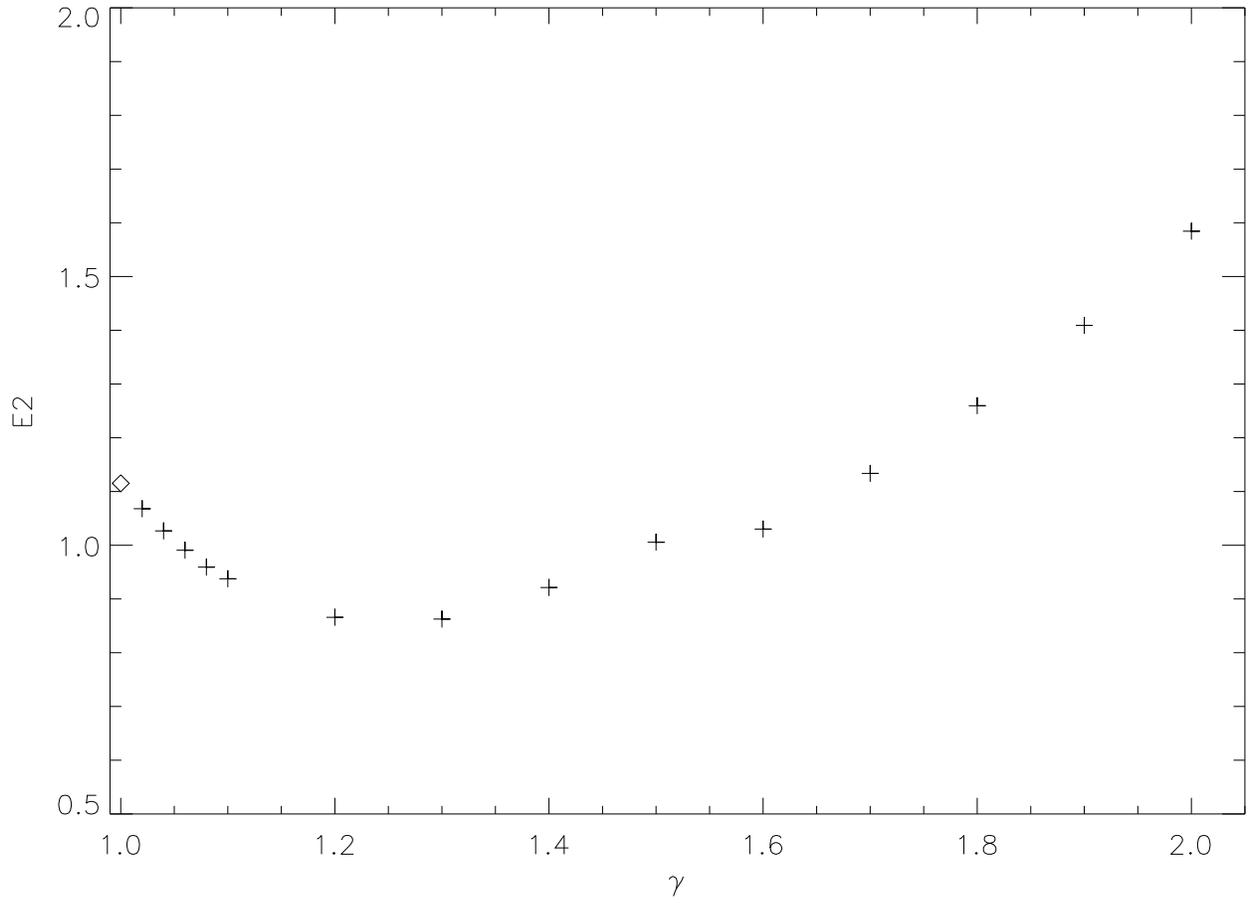}\caption{The dependence of $E_2$ on $\gamma$.
Spherical clusters are considered and the mass is taken to be
$10^{14}h^{-1}M_\odot$.
\label{fig4}}
\end{figure}

\begin{figure}
\plotone{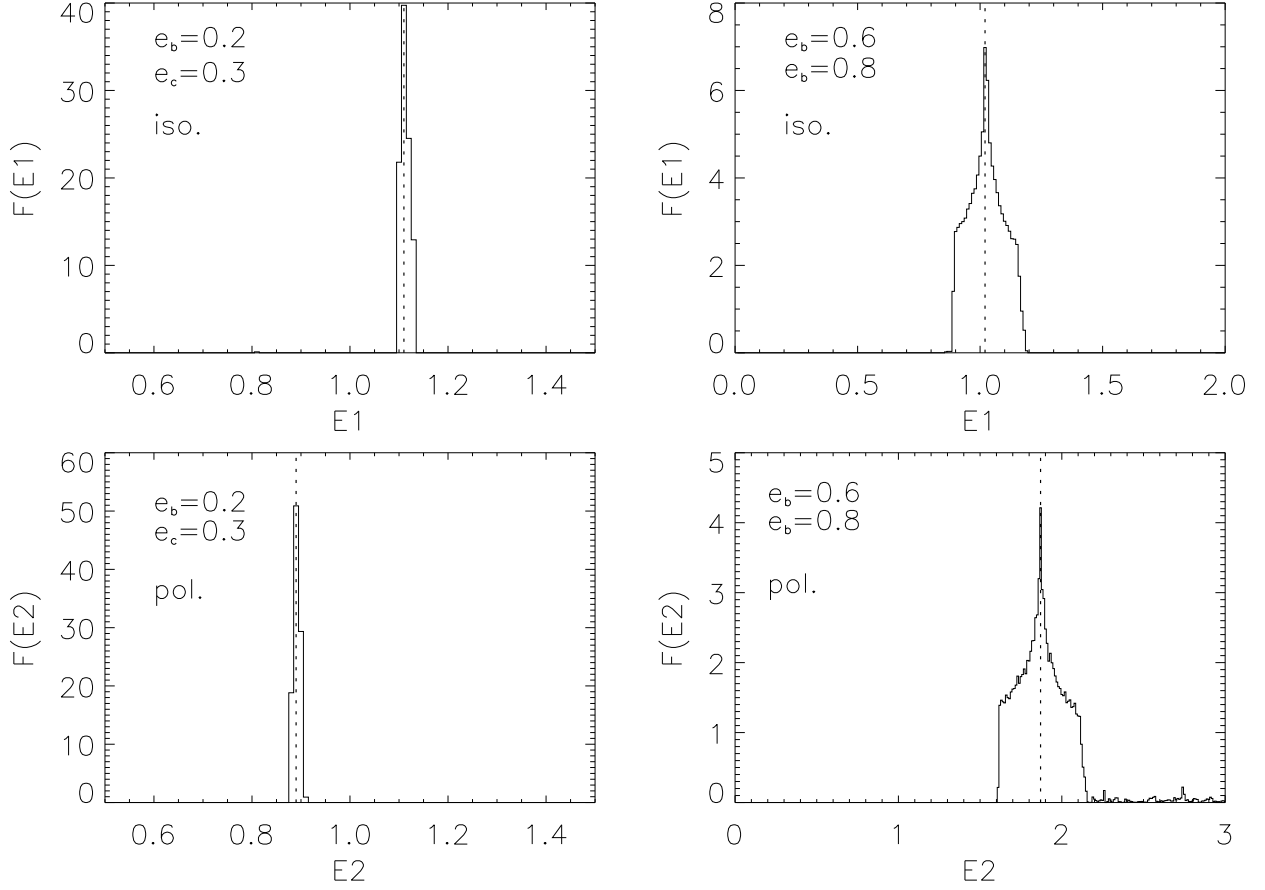}\caption{Conditional probabilities $F(E_1|c_e,
c/b, c/a)$ and $F(E_2|c_e, c/b, c/a)$ for clusters with
$M=10^{14}h^{-1}M_{\odot}$ and $z=0.1$. The top and bottom
panels show the isothermal and polytropic cases, respectively.
The vertical lines indicate the peak values of $E_1(E_2)$.
\label{fig5}}
\end{figure}

\begin{figure}
\plotone{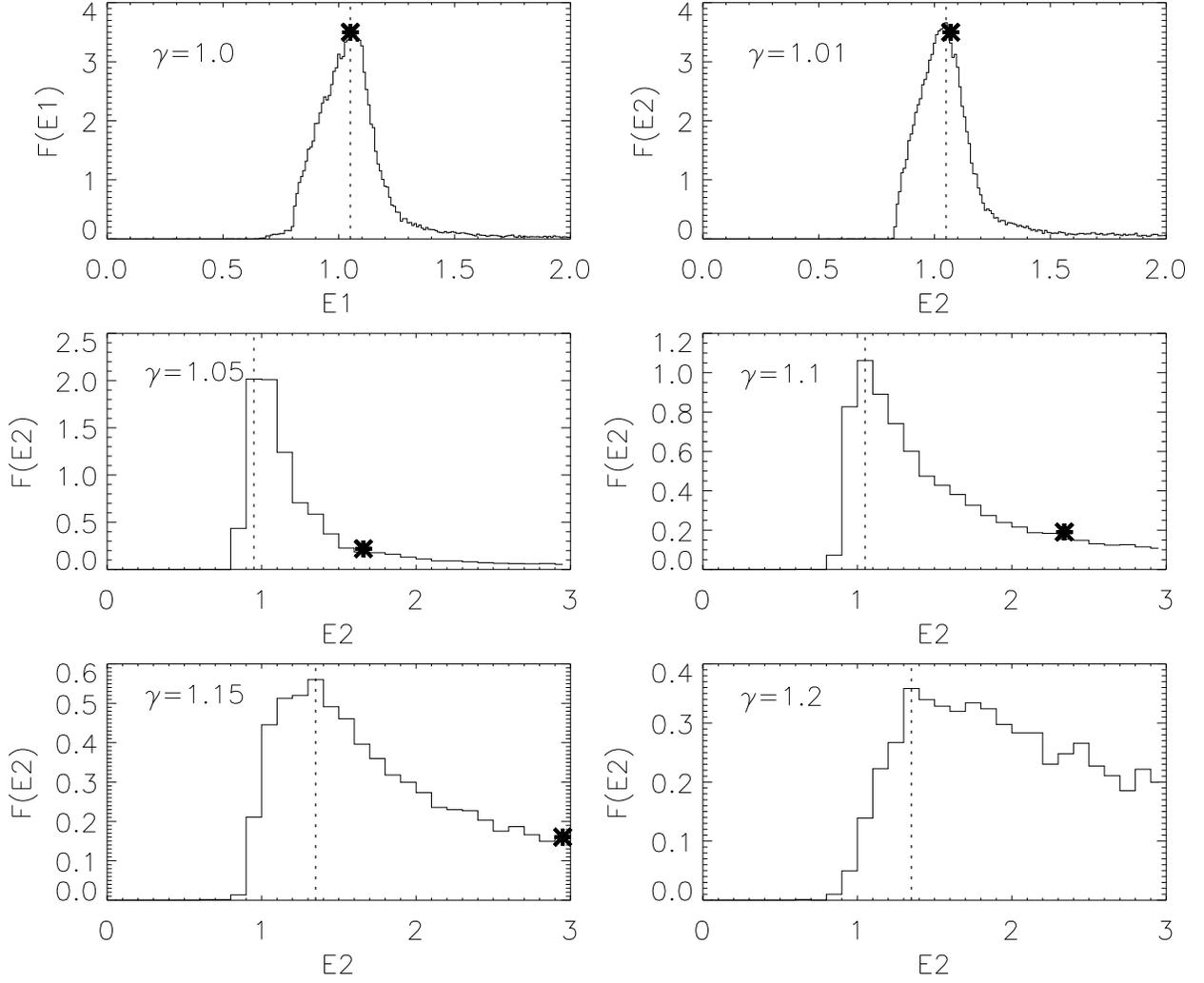}\caption{Probability functions $F(E_1|{M,z})$ (top
left panel) and $F(E_2|{M,z})$ with $M=10^{14}h^{-1}M_{\odot}$ and
redshift $z=0.1$. In each panel, the peak value of $E_1$ ($E_2$) is
indicated by the vertical line. The symbols show the
average values of $E_1(E_2)$. \label{fig6}}
\end{figure}

\begin{figure}
\plotone{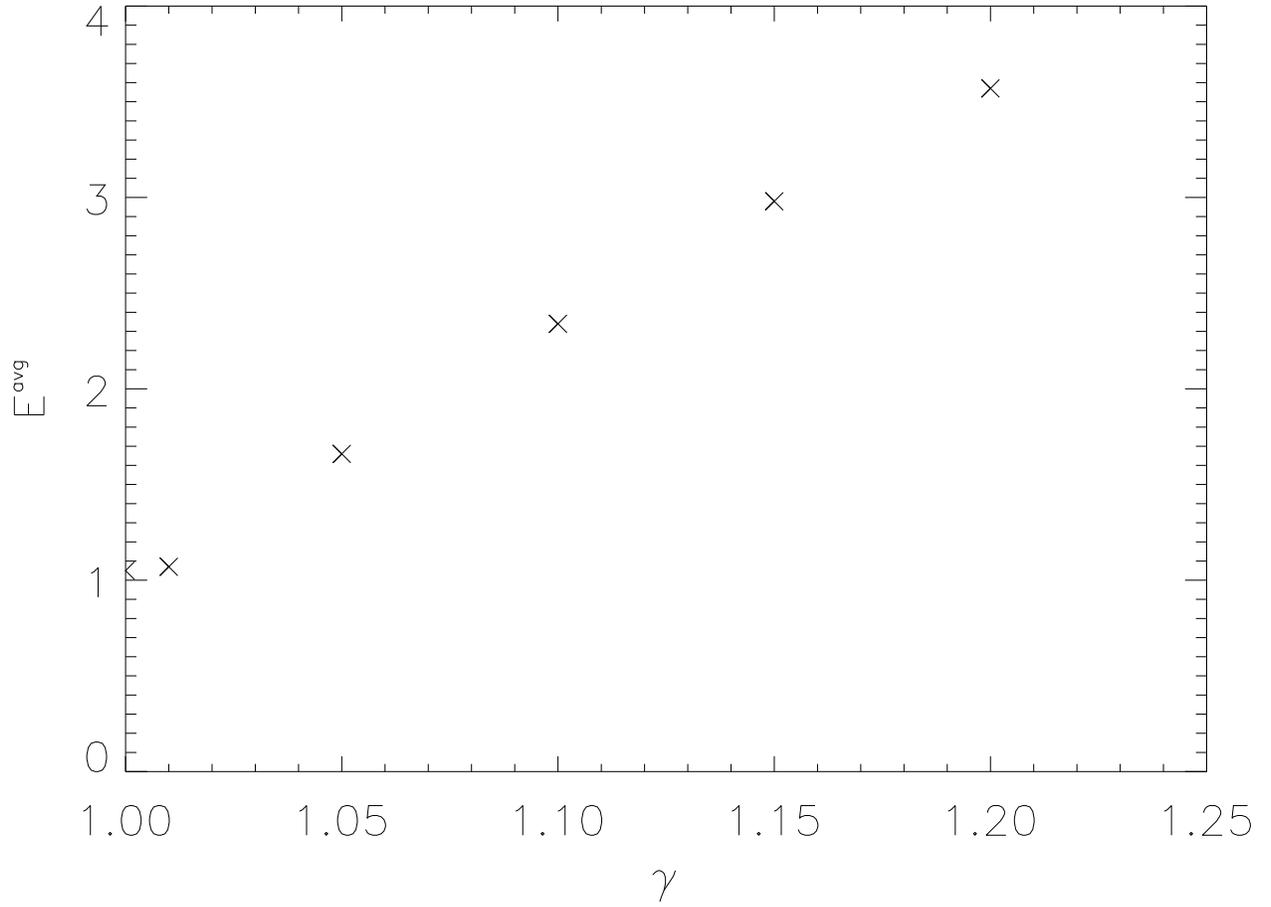}\caption{The dependence of the $E_2^{ave}$ on the
polytropic index $\gamma$. We take $M=10^{14}h^{-1}M_{\odot}$ and
redshift $z=0.1$.
\label{fig7}}
\end{figure}

\begin{figure}
\plotone{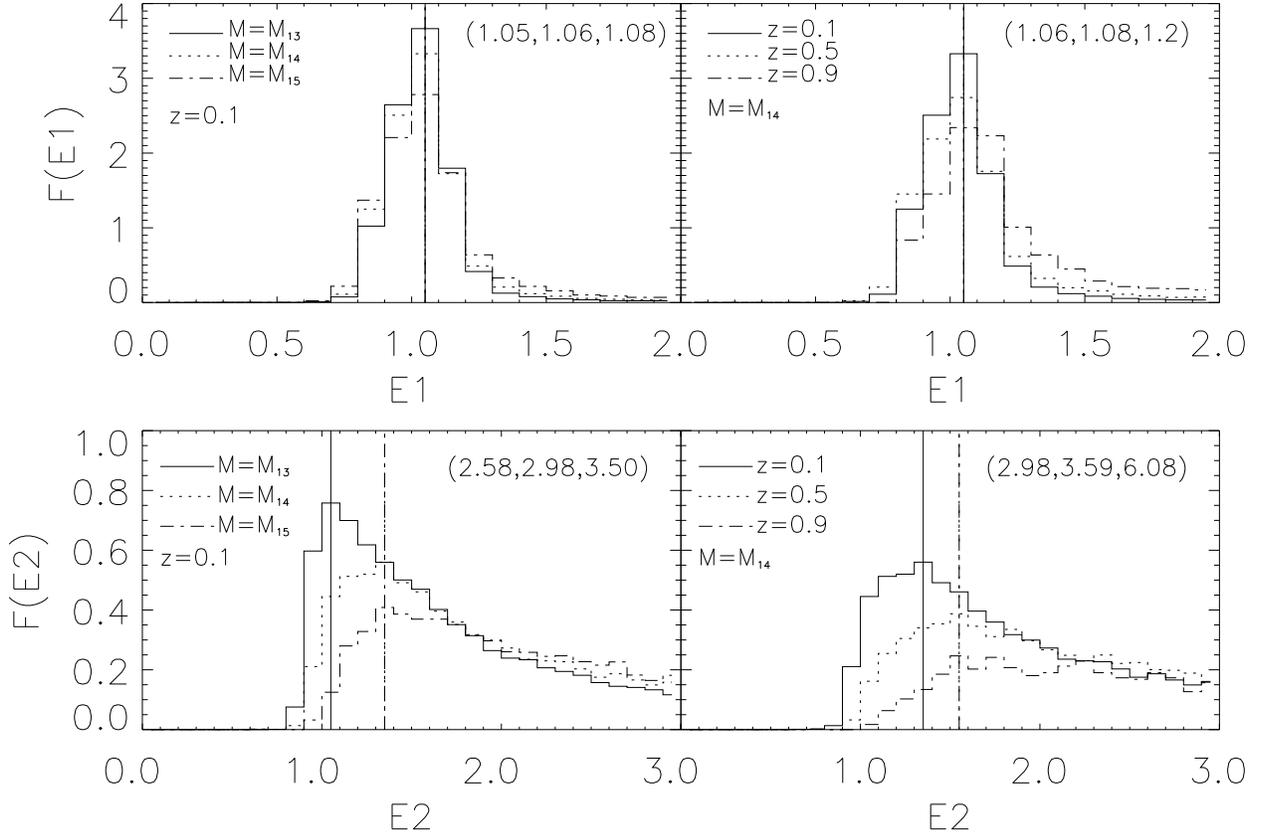}\caption{Probability distributions $F(E_1|{M,z})$
and $F(E_2|{M,z})$. The top and bottom panels
are for the isothermal and polytropic cases, respectively.
The left ones show the dependence of the probabilities on $M$,
and the right ones demonstrate the change of the probabilities
with the change of $z$.
In each panel, the numbers listed in the parentheses are
the average values of $E_1$ ($E_2$) for the three cases, respectively.
The vertical lines show the peak values of $E_1(E_2)$.
\label{fig8}}
\end{figure}

\begin{figure}
\plotone{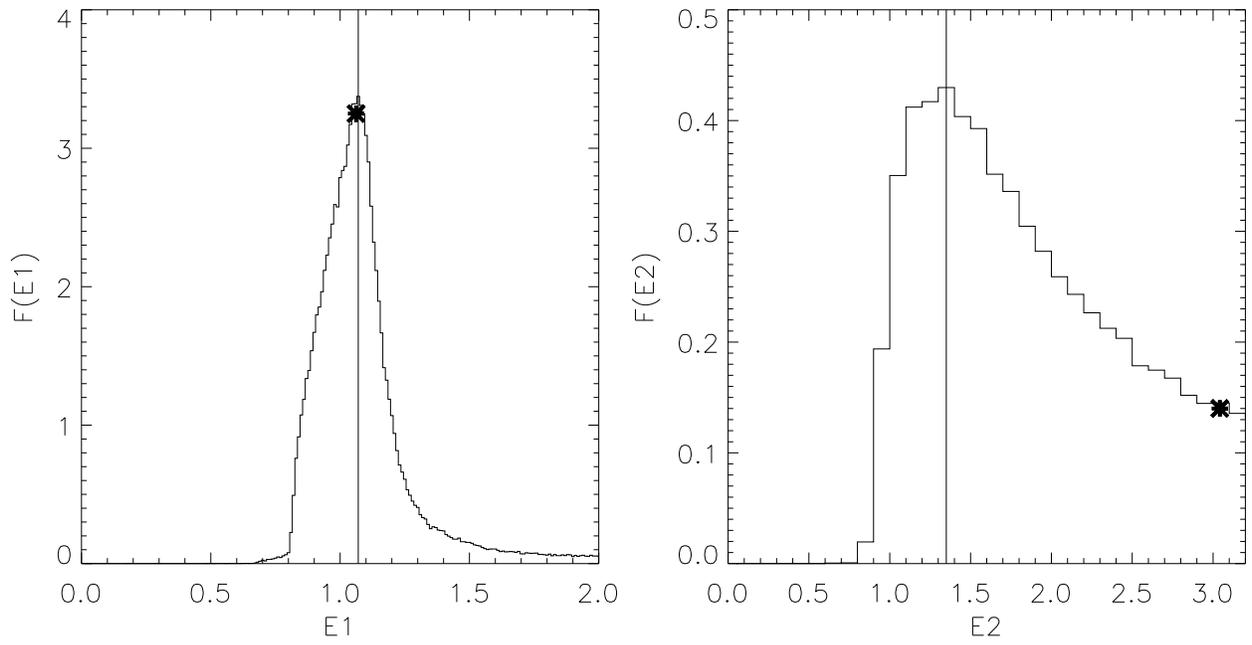}\caption{Probability functions $F(E_1)$ and $F(E_2)$
for mass-limited cluster samples with
$M_\mathrm{lim}=10^{13}h^{-1}M_{\odot}$,
The vertical lines denote the peak values of $E_1$ and
$E_2$. The symbols show the average values of $E_1$ and $E_2$. \label{fig9}}
\end{figure}

\begin{figure}
\plotone{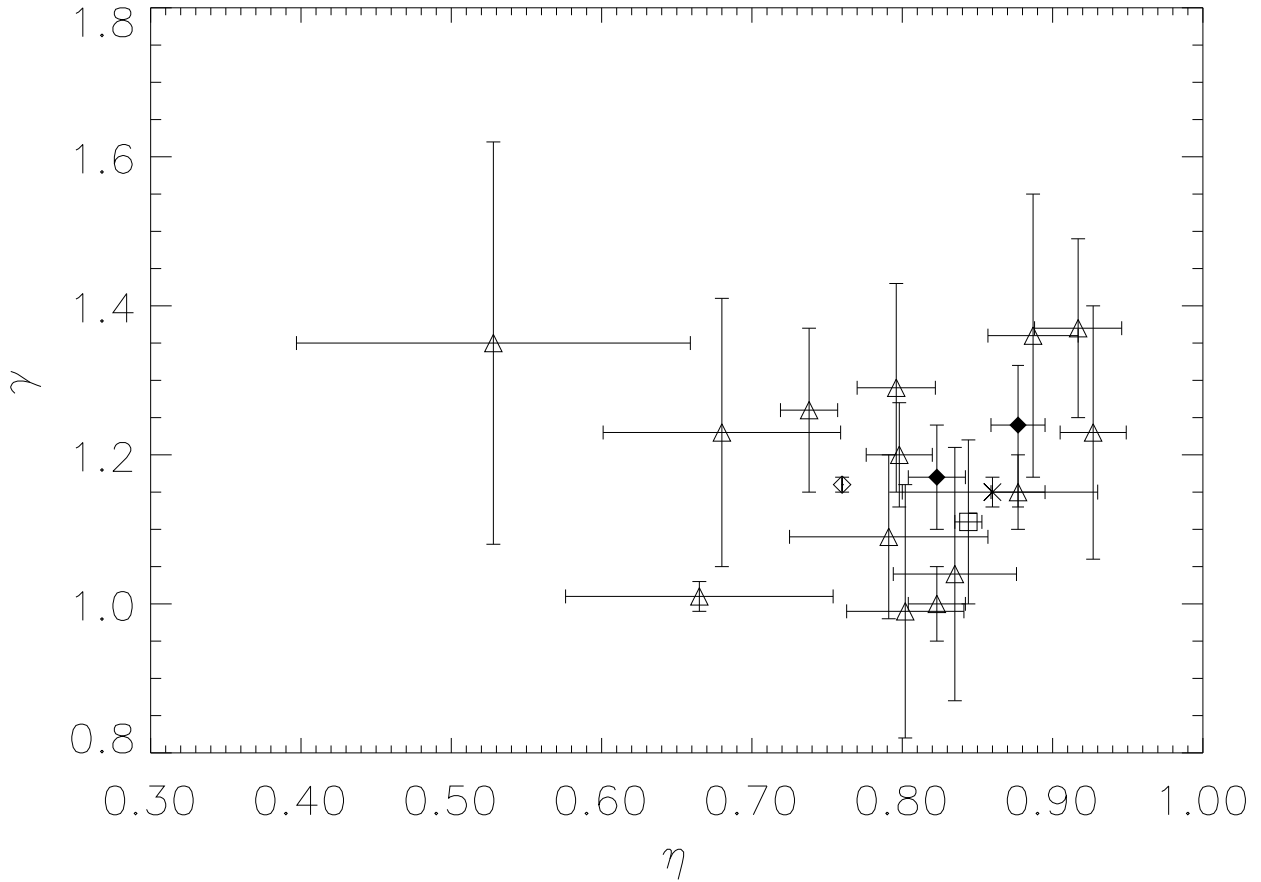}\caption{Scatter plot of $\gamma$ versus the axial ratio
of the X-ray profile $\eta$ for $19$ clusters. The triangles
are from Mohr et al. (1995) ($\eta$) and Finoguenov et al. (2001)
($\gamma$). The filled diamonds are from Mohr et al. (1995) ($\eta$)
and Markevitch et al. (1999) ($\gamma$). The square is from
De Filippis et al. (2005) ($\eta$) and Finoguenov et al. (2001)
($\gamma$). The cross is from
De Filippis et al. (2005) ($\eta$) and Pratt et al. (2005) ($\gamma$).
The open diamond is from Belsole et al. (2005) ($\eta$ and $\gamma$).
\label{fig10}}
\end{figure}

\begin{figure}
\plotone{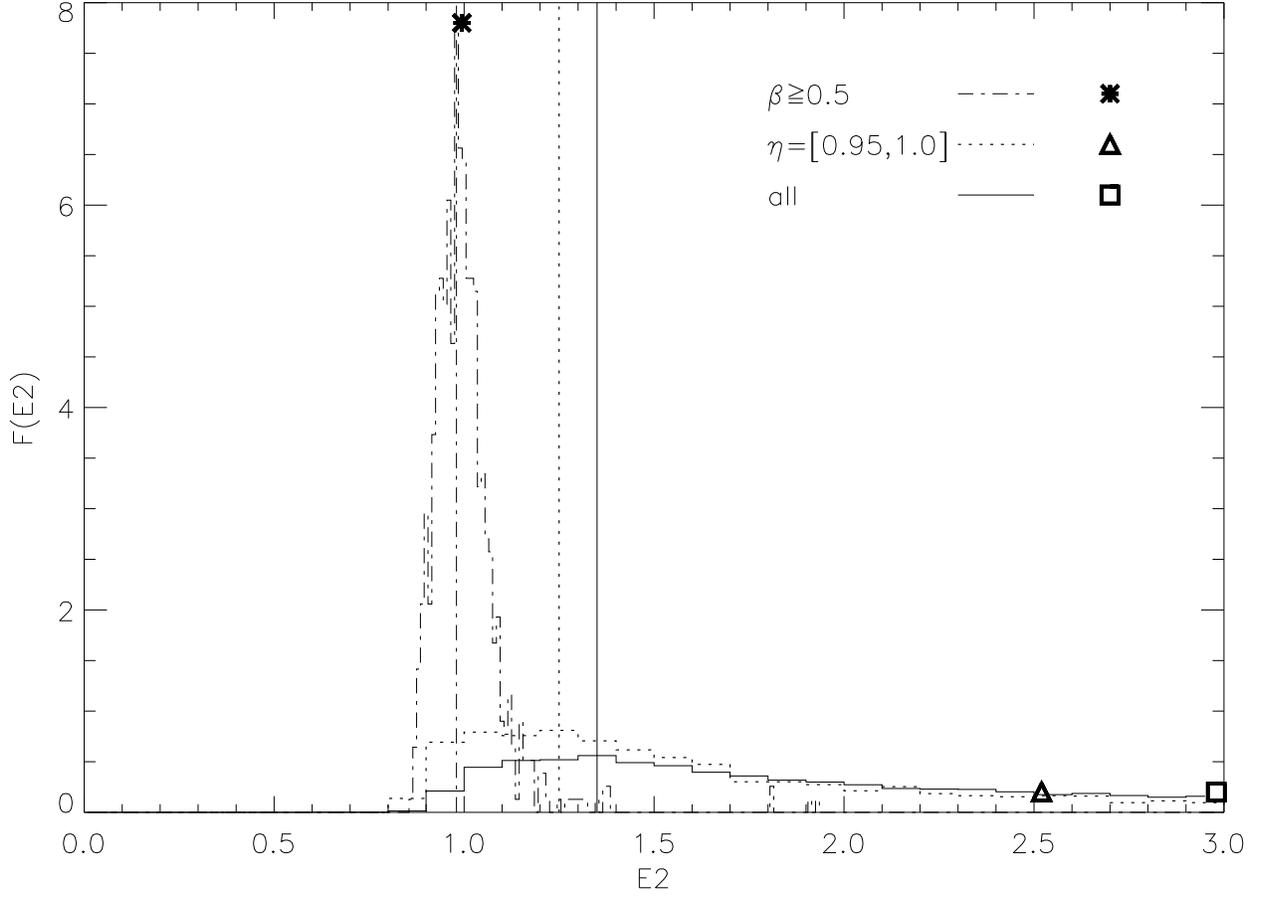}\caption{The distributions of $E_2$.
The considered sample of clusters is
$M=10^{14}h^{-1}\hbox{ M}_{\odot}$
and redshift $z=0.1$. The polytropic index $\gamma=1.15$.
The solid, dotted, and dash-dotted lines are
for the distributions of the full sample, the subsample
with $\eta\ge 0.95$ and the subsample with $\beta\ge 0.5$.
The positions of the vertical lines and the symbols
indicate the peak values and the average values of $E_2$
of the corresponding distributions.
\label{fig11}}
\end{figure}

\begin{figure}
\plotone{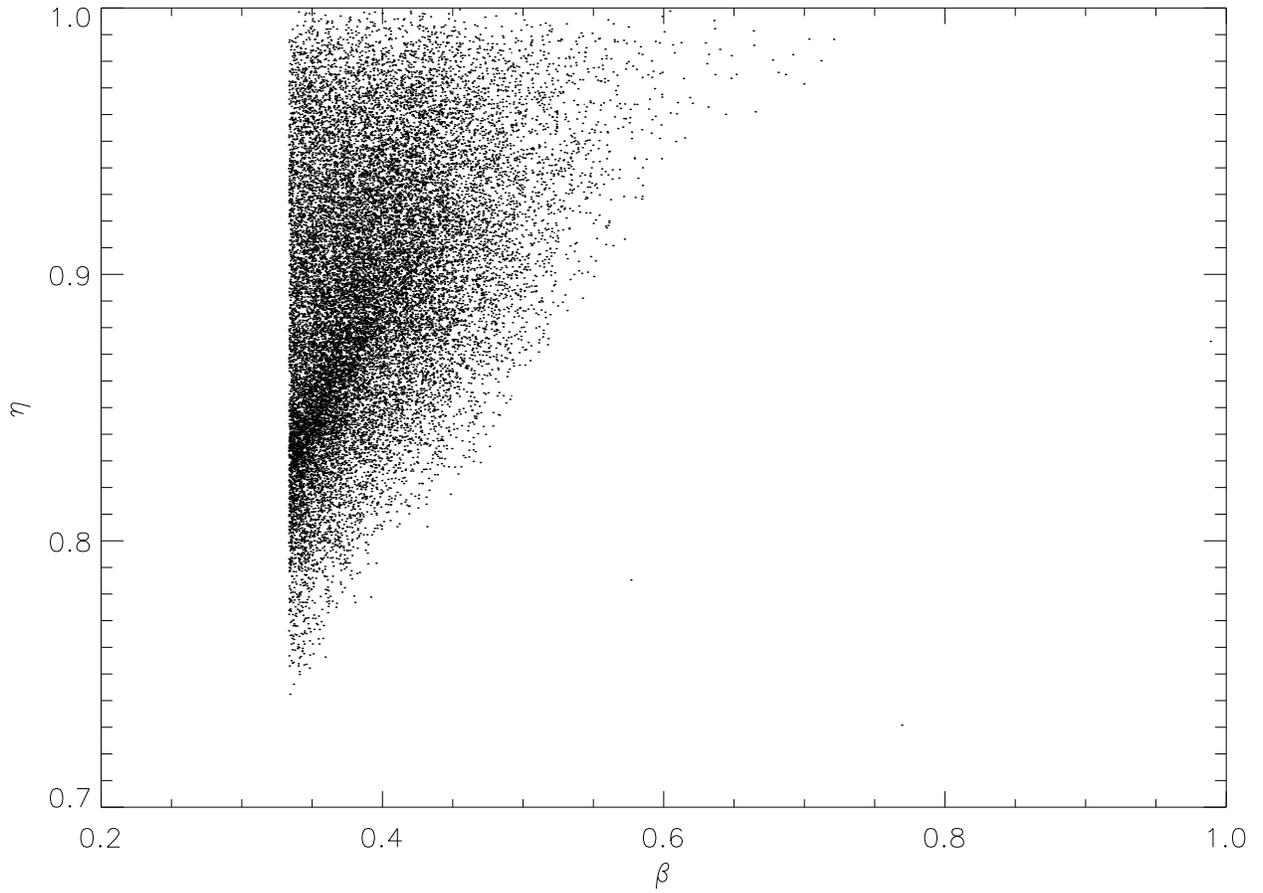}\caption{The scatter plot of $\beta$ vs.
$\eta$ for the sample of clusters with
$M=10^{14}h^{-1}\hbox{ M}_{\odot}$ and redshift $z=0.1$.
The polytropic index $\gamma=1.15$.
\label{fig12}}
\end{figure}

\begin{figure}
\plotone{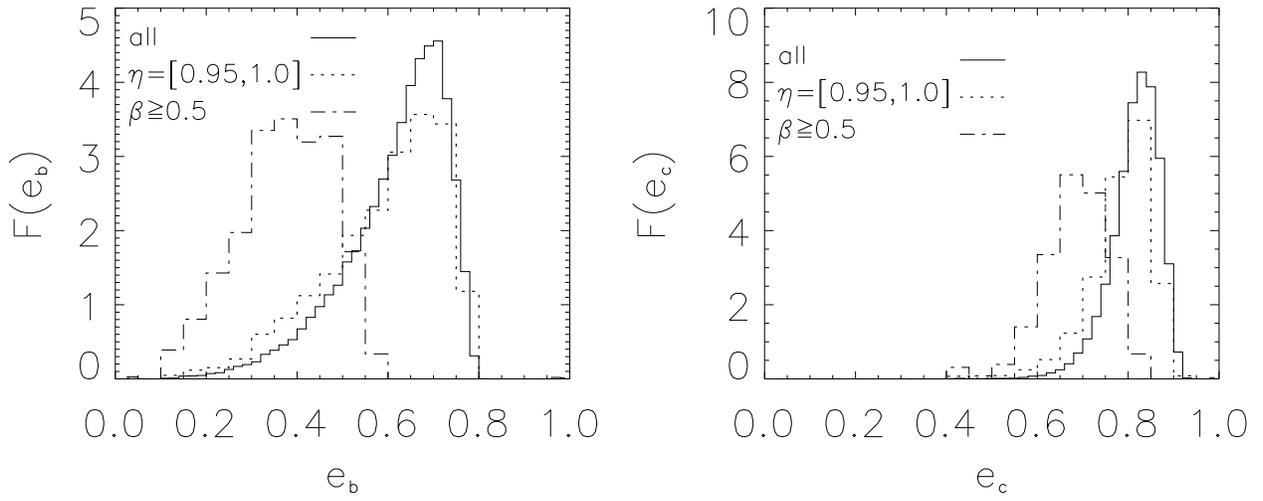}\caption{The distributions of $e_b$
(left panel) and $e_c$ (right panel) for the samples
considered in Figure 11.
\label{fig13}}
\end{figure}

\begin{figure}
\plotone{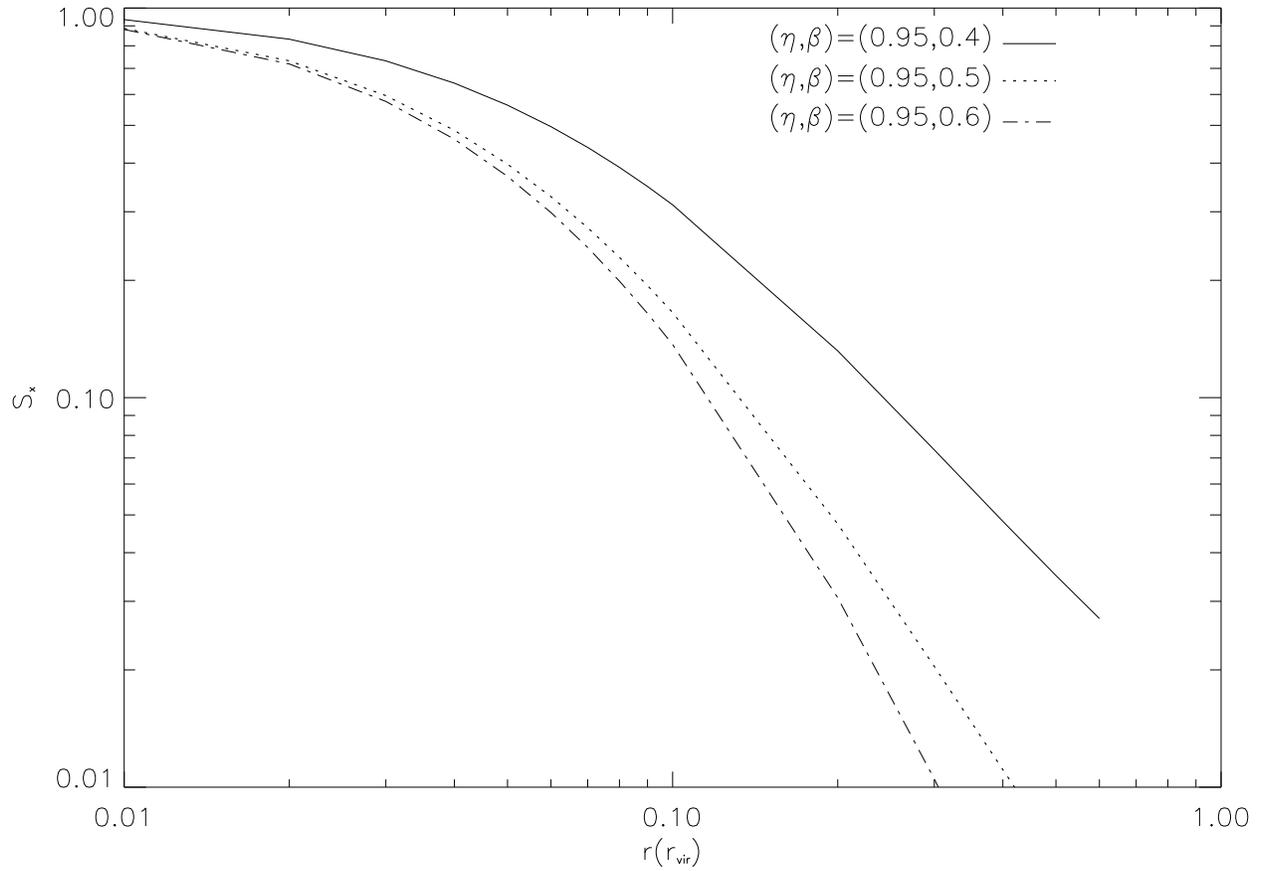}\caption{The circularized radial
profile of X-ray emission for three specific clusters
selected from the sample shown in Figure 12.
For all the three clusters, we have $\eta=0.95$.
The fitted $\beta$ values are $0.4,0.5,$ and $0.6$,
respectively.
\label{fig14}}
\end{figure}

\end{document}